\documentclass[a4paper,11pt]{article}
\usepackage{cite}

\newcommand{\email}[1]{\footnote{email: #1} }

\usepackage{amsmath, amssymb}
\usepackage[margin=1in]{geometry}
\usepackage{booktabs}
\usepackage{xcolor}
\usepackage{slashed}
\usepackage{tikz}
\usetikzlibrary{decorations.pathmorphing,decorations.markings}

\newcommand{\Sf}{S}
\newcommand{\SfI}{S^{-1}}
\newcommand{\dL}{\displaystyle\int\!\frac{d^d\ell}{(2\pi)^d}}
\newcommand{\Tr}{{\rm tr} }
\newcommand{\Dp}{{D^P}}
\newcommand{\Gf}{\Gamma_F}
\newcommand{\Gp}{\Gamma_P}

\title{Gauge-Invariant Off-Shell Mass}
\author{Kang-Sin Choi\,\email{kangsin@ewha.ac.kr}   
and Hyeseon Im\,\email{lisa\_im0814@ewha.ac.kr}
\\ \it \normalsize $^*$ Scranton Honors Program, Ewha Womans University, Seoul 03760, Korea \\ \it \normalsize $^*$ Institute of Mathematical Sciences, Ewha Womans University, Seoul 03760, Korea\\ \it \normalsize $^\dagger$ Department of Physics, Ewha Womans University, Seoul 03760, Korea}

\date{}

\begin{document}
\maketitle

\begin{abstract}
We define a gauge-invariant and renormalized off-shell mass function in quantum field theory. The conventional self-energy is gauge-dependent off-shell. We generalize the self-energy, together with the corresponding vertex function, to be gauge-invariant and process-independent. To do this, we extend the pinch technique to an arbitrarily long fermion line. The Ward--Takahashi identity acting locally at each vertex cancels the gauge-dependent piece exactly, leaving the Feynman-gauge self-energy. On-shell renormalization then yields a mass function that is gauge-invariant, scheme-independent and equal to the physical mass at the pole; the same self-energy also yields an infrared-finite scalar mass function directly comparable with Schwinger--Dyson and lattice determinations. The cancellation is local to the internal fermion line and holds entirely off-shell, so the self-energy is well-defined on each internal segment of an amplitude with arbitrarily many external photons; we demonstrate this explicitly by computing the minimal off-shell Compton amplitude.
\end{abstract}


\section{Introduction}
\label{sec:intro}

The concept of mass in quantum field theory is fundamentally different from that in classical mechanics. The bare mass $m_B$ is a parameter of the Lagrangian, defined in the absence of further interactions, and is not directly measurable quantum mechanically. When the fermion interacts, loop corrections shift it to $m_B + \Sigma(q)$, where $\Sigma(q)$ is the self-energy, and the dressed propagator is $\Sf(q) = i/(\slashed{q}-m_B-\Sigma(q))$. The observed mass is determined by the interference between the tree-level amplitude and the loop corrections in a scattering process, so mass and propagator are inseparable.

At the pole $\slashed{q} = m$, the dressed propagator has a simple pole, and the pole mass $m$ coincides with the asymptotic mass appearing in the Lehmann--Symanzik--Zimmermann reduction formula. The pole mass is gauge-independent, as demonstrated by the Nielsen identity~\cite{Nielsen:1975, Breckenridge:1994gs,Gambino:1999ai}.

Our goal is to define the off-shell mass, that is, the mass function evaluated at arbitrary virtuality $\slashed{q} \neq m$ \cite{Choi:2026yxc}. The off-shell propagator enters every amplitude as an internal line, and the momentum dependence of the mass governs running. At large spacelike $q^2$ the off-shell mass function is the running mass in the same sense that the effective charge $\alpha(q^2)$~\cite{Cornwall:1982,Grunberg:1984,Watson:1997} is the running coupling: a gauge-invariant, process-independent function of the physical momentum, of which the $\bar{m}(\mu)$ of the renormalization group is the scheme-dependent tracker, the two related by finite matching \cite{Choi:2024cbs,Choi:2024hkd}, as made precise in Section~\ref{sec:mass}.

The question is whether this off-shell mass function can be defined in a gauge-invariant way. The difficulty is structural. In $R_\xi$ gauges the boson propagator is gauge-parameter dependent, with the $\xi$-dependence carried by the longitudinal term $i (1-\xi) \ell_\mu\ell_\nu/\ell^4$, which attaches to the fermion line at the two endpoints of the virtual photon, so the $(1-\xi)$ contamination of any one-loop fermion correction is distributed across the global topology of the diagram rather than residing on any single segment of the fermion line. The total $S$-matrix is gauge-invariant~\cite{Abers:1973qs}, so by the Ward--Takahashi identity (WTI)~\cite{Ward:1950,Takahashi:1957} the gauge-dependent pieces from self-energy, vertex and box topologies cancel when summed, but at first sight the cancellation appears to require combining contributions from non-adjacent regions of the diagram, in a way that depends on the global topology.

The key observation of this paper is that, despite this global appearance, the cancellation is {\em local to the segment}. The gauge-dependent part of a self-energy on one interval of the fermion line is removed by the WTI firing at the two vertices bounding that interval, and by nothing else. The firing produces an inverse propagator $\SfI$ that eats its neighboring propagator, and the telescoping stays within the segment bounded by the external photon vertices. The global structure of the diagram enters only through boundary terms pushed outward onto neighboring segments, never back onto the interval, so the gauge invariance of the self-energy is a property of its own interval. The cancellation requires neither on-shell external fermions nor the Dirac equation.

This segment-local mechanism provides a natural decomposition of the gauge-dependent pieces. When the WTI fires at a vertex adjacent to an internal fermion line, it produces a difference of propagators $\Sf(q) - \Sf(q-\ell)$ that modifies only the internal line while leaving the external legs as spectators. These are the propagator-like pinch contributions; absorbing them into the conventional self-energy defines a generalized gauge-invariant self-energy $\hat\Sigma(q)$. The remaining pieces, where the WTI shifts an external-leg propagator, are vertex-like and define the modified vertex $\hat\Gamma$. All gauge-dependent pieces are proportional to $(1-\xi)$ and cancel among themselves, so $\hat\Sigma(q) = \Sigma_\xi(q)\big|_{\xi=1}$. The renormalized mass function $m(q) = m + \hat\Sigma^{\text{ren}}(q)$ is then gauge-invariant at every momentum, equals the physical mass at the pole ($m(m) = m$) and has no residual scheme or regularization dependence when expressed in on-shell parameters.

This approach builds on the pinch technique (PT). The standard PT~\cite{Cornwall:1982,Cornwall:Papavassiliou:1989,Papavassiliou:1990,Degrassi:Sirlin:1992,Papavassiliou:1995,Binosi:Papavassiliou:2002,Binosi:Papavassiliou:2004,Binosi:Papavassiliou:2009,Cornwall:Papavassiliou:Binosi:2011} constructs gauge-invariant off-shell Green's functions by extracting them from on-shell $S$-matrix elements. The Dirac equation  is used at every external pinch vertex, and ``off-shell'' in the PT literature refers to the momentum argument of the extracted Green's function (e.g., $\hat\Pi(q^2)$ at arbitrary $q^2$), not to the external test fermions, which remain on-shell. The identification of ``propagator-like'' pinch parts relies on their Dirac structure after the on-shell projection, which makes process independence a nontrivial property to be established rather than one that is manifest; Binosi and Papavassiliou~\cite{Binosi:Papavassiliou:2004} confirmed it explicitly, showing that the gluon self-energy extracted from quark external states matches the one from gluon external states. The all-orders extension in QCD requires the Batalin--Vilkovisky (BV) formalism~\cite{Binosi:Papavassiliou:2002:BV} because the Slavnov--Taylor identity~\cite{Taylor:1971,Slavnov:1972} involves ghosts.

In both established formulations of the PT these boundary terms are discarded rather than kept: in the $S$-matrix PT they vanish on the external spinors through the Dirac equation, and in the intrinsic PT~\cite{Cornwall:Papavassiliou:1989} they are isolated as the parts proportional to the external inverse propagators and dropped~\cite{Hashimoto:Kodaira:Yasui:Sasaki:1994,Binosi:Papavassiliou:2002:BV}, again because they would vanish on-shell. In neither case are the external fermions off-shell. The present construction instead removes the gauge-dependent part by the segment-local WTI on the intermediate line, an algebraic identity that needs no on-shell projection, so the ``propagator-like'' parts are fixed intrinsically rather than read off after one. We show this in the minimal Compton amplitude (Section~\ref{sec:explicit}) with the external legs off-shell, the residual boundary terms canceling against the external-leg self-energies and the crossed ($u$-channel) amplitude. Because the self-energy depends only on the internal momentum, the general-$n$ and all-orders arguments (Sections~\ref{sec:general_n}--\ref{sec:allorders}) may instead place the outermost legs on-shell, where the boundary terms vanish at once, with the same result. Process-independence, verified in the standard PT by explicit calculation across external states~\cite{Binosi:Papavassiliou:2004}, here follows from the locality of the segment.

The equivalence between the PT and the background field method (BFM) in the Feynman gauge~\cite{Denner:Dittmaier:Weiglein:1994,Hashimoto:Kodaira:Yasui:Sasaki:1994,Papavassiliou:1995,Binosi:Papavassiliou:2002:BV} guarantees that gauge-invariant Green's functions exist off-shell, since the BFM does not reference on-shell spinors. However, the BFM achieves this by modifying the gauge-fixing procedure (splitting the gauge field into background and quantum parts), not by analysis of the original $R_\xi$ gauge theory. One could in principle extract a gauge-invariant $m(q)$ from the BFM fermion propagator; to our knowledge this step has not been taken, the emphasis of the BFM program having naturally been on the gauge sector (gluon propagator, three-gluon vertex, ghost sector), where the most pressing dynamical questions lie.

On the dynamical side, Cornwall~\cite{Cornwall:1982} showed that the gluon propagator acquires a momentum-dependent mass $m^2(q^2)$ through the Schwinger mechanism~\cite{Schwinger:1962} in the PT framework, and subsequent work~\cite{Aguilar:Papavassiliou:2006,Aguilar:Binosi:Papavassiliou:2010,Binosi:Papavassiliou:2008:trunc} computed this mass function from Schwinger--Dyson equations (SDEs) in the PT-BFM truncation scheme. These works concern the gluon sector, where the Schwinger mechanism operates. The corresponding treatment of the fermion mass function within the same gauge-invariant framework is a natural next step that, to our knowledge, has not yet been carried out. The Schwinger--Dyson community~\cite{Roberts:Schmidt:2020} routinely computes the quark mass function $m(q^2)$, demonstrating dynamical chiral symmetry breaking ($m(0)\sim 300$~MeV from nearly massless current quarks), and lattice QCD determines the same mass function directly~\cite{Zhang:2004,Bowman:2005}; both are carried out in a fixed gauge, typically Landau gauge, and the resulting $m(q^2)$ is accordingly gauge-dependent, its gauge covariance governed by the Landau--Khalatnikov--Fradkin transformations~\cite{Landau:Khalatnikov:1955}, as those works themselves make explicit.

The gauge-invariant fermion self-energy $\hat\Sigma$ is therefore available in principle; Binosi and Papavassiliou~\cite{Binosi:Papavassiliou:2002} computed it explicitly at two loops in both QED and QCD, where it served to demonstrate the consistency of the framework. The step taken here is complementary: we promote the mass function $m(q) = m + \hat\Sigma(q)$ to the primary object and develop its physical content, a gauge-invariant mass at every virtuality. Meanwhile, the existing gauge-invariant mass definitions each cover only part of the momentum range. The pole mass $m$ is a single number~\cite{Nielsen:1975,Tarrach:1981}, in QCD moreover infrared-sensitive at the nonperturbative level~\cite{Bigi:1994,Beneke:Braun:1994}; for unstable particles it is promoted to the gauge-invariant complex pole of the propagator~\cite{Sirlin:1991,Willenbrock:Valencia:1991,Stuart:1991,Kniehl:Sirlin:2008}, with the pinch technique supplying the gauge-invariant resummed off-shell self-energy on which that definition rests~\cite{Papavassiliou:Pilaftsis:1995,Papavassiliou:Pilaftsis:1996,Papavassiliou:Pilaftsis:1996b}, but this is still a single point in the complex plane rather than a function of virtuality, and the on-shell mass away from the pole carries an unbounded gauge dependence~\cite{Passera:Sirlin:1998}. The $\overline{\text{MS}}$ running mass $\bar{m}(\mu)$ is gauge-invariant but scheme-dependent and does not coincide with the mass in the dressed propagator~\cite{Tarrach:1981,Kronfeld:1998}. The full momentum-dependent mass function, the quantity that actually governs fermion propagation at arbitrary virtuality, has remained without a gauge-invariant, scheme-independent definition.

The gauge-invariant mass function was introduced in~\cite{Choi:2026yxc}; in this work, we carry out the underlying construction in detail. We use the $s$-channel Compton amplitude as the minimal process for defining $\hat\Sigma$, with the $u$-channel needed only for boundary-term cancellation; the resulting $\hat\Sigma$ is process-independent and coincides with the Feynman-gauge self-energy. We renormalize on shell, identify the infrared content of the wave-function subtraction and extract from the same self-energy an infrared-finite, renormalization-group invariant scalar mass function directly comparable with Schwinger--Dyson and lattice determinations. We then extend the construction to a fermion line with an arbitrary number $n+1$ of external photon vertices, showing that the gauge-dependent contributions from topologies with $k$ spectator photons cancel by a telescoping mechanism across adjacent topologies $k-1$, $k$, $k+1$. The construction carries over to non-Abelian gauge theory, where pinch decomposition of the three-gluon vertex and the ghost sector supply on gluon segments what the Ward--Takahashi identity supplies on fermion segments, as we illustrate with the off-shell quark-gluon Compton amplitude. The same segment-local mechanism continues to all orders by induction on the loop order, the exact WTI replacing the tree-level identity.

\section{Gauge Dependence and the Pinch Technique}
\label{sec:pinch}

We work in a general linear covariant gauge theory in which a fermion couples to a gauge boson through the vertex $-ig t^a\gamma^\mu$. For QED, the gauge group is $U(1)$, we drop $t^a$, and $g$ is identified with the electromagnetic coupling $e$. For ``QCD'' with the gauge group $SU(N)$, $t^a (a=1,\ldots,N^2-1)$ are the generators of the fundamental representation with $\Tr(t^at^b)=\tfrac12\delta^{ab}$, and $g$ is the coupling. The fermion propagator is 
\begin{equation}
  \Sf(p) = i \frac{\slashed p + m}{p^2 - m^2 + i\epsilon}, \qquad
  \SfI(p) = -i(\slashed p - m), \qquad
  \Sf(p) \SfI(p) = \mathbf{1},
  \label{eq:PS_fermion}
\end{equation}
so that the Ward--Takahashi identity at tree level reads
\begin{equation}
  \slashed\ell = i\bigl[\SfI(k') - \SfI(k)\bigr]
  \qquad\text{with}\qquad \ell = k' - k,
  \label{eq:WTI_tree}
\end{equation}
where $\ell$ is the photon momentum flowing into the vertex and $k$, $k'$ are the incoming and outgoing fermion momenta. The gauge-boson propagator in $R_\xi$ gauge has the same Lorentz structure in QED and QCD,
\begin{equation}
  D^{ab}_{\mu\nu}(\ell) = \delta^{ab} D_{\mu\nu}(\ell), \qquad
  D_{\mu\nu}(\ell) = \frac{-i}{\ell^2}\left[g_{\mu\nu}
  - (1-\xi)\frac{\ell_\mu\ell_\nu}{\ell^2}\right].
\end{equation}

\subsection{The fermion self-energy}

The full fermion propagator is $S(q)=i/(\slashed q-m_B-\Sigma(q))$, with $\Sigma(q)$ the one-particle-irreducible self-energy, the sum of all 1PI insertions on the internal fermion line. This $\Sigma$ and the mass function read from the dressed inverse propagator are what we aim to define gauge-invariantly. At one loop,
\begin{equation}
  -i\Sigma(q)
  = (-ig)^2 C_2\dL \gamma^\mu \Sf(q-\ell) \gamma^\nu 
    D_{\mu\nu}(\ell),
  \label{eq:SE}
\end{equation}
where $D_{\mu\nu}(\ell)$ is the full gauge-boson propagator in $R_\xi$ gauge and the color factor $t^at^a = C_2 \mathbf{1}_F$ has been contracted at the two internal vertices. We write $C_2$ for the Casimir generated by $t^at^a$, with $C_2 = 1$ in QED and $C_2 = C_F = (N^2-1)/(2N)$ in QCD. The genuinely non-abelian content of the theory (the three-gauge-boson self-coupling, the ghost sector and the consequences for diagrams with external gauge bosons carrying color) does not enter the one-loop fermion self-energy with external photon vertices and is treated separately in Section~\ref{sec:qcd}. 

We decompose the gauge-boson propagator into its Feynman-gauge part and the ``pinch'' part,
\begin{equation}
  D_{\mu\nu}(\ell) = D^F_{\mu\nu}(\ell) + D^P_{\mu\nu}(\ell),
  \label{eq:Ddecomp}
\end{equation}
where
\begin{equation}
  D^F_{\mu\nu}(\ell) =-i \frac{g_{\mu\nu}}{\ell^2},
  \qquad
  D^P_{\mu\nu}(\ell) = i (1-\xi) \frac{\ell_\mu\ell_\nu}{\ell^4}.
  \label{eq:DFandDP}
\end{equation}

Inserting the decomposition \eqref{eq:Ddecomp} into \eqref{eq:SE}, the self-energy splits accordingly
\begin{equation}
  \Sigma(q)
  = \Sigma_F(q) + \Sigma_P(q),
\end{equation}
where $\Sigma_F(q)$ and $\Sigma_P(q)$ are the contributions from $D^F$ and $D^P$, respectively. Since $D^P\propto(1-\xi)$, the gauge dependence of the self-energy resides entirely in $\Sigma_P$. Contracting the $\ell^\mu\ell^\nu$ of $D^P$ at the two vertices brings down $\slashed\ell\,\Sf(q-\ell)\,\slashed\ell$. Applying the tree Ward--Takahashi identity~\eqref{eq:WTI_tree} at each $\slashed\ell$ insertion leaves every resulting term with an overall factor $\SfI(q)\propto(\slashed{q}-m)$, so that
\begin{equation} \label{SigmaP}
  \Sigma_P(q) = (1-\xi)\,(\slashed{q}-m)\,\Omega(q),
\end{equation}
where $\Omega(q)$ is a gauge-parameter-independent, Dirac-valued one-loop coefficient whose explicit form is given in the appendix. Only the overall factor $(\slashed{q}-m)$ enters what follows. It vanishes on-shell $\slashed{q}=m$, so the self-energy, and hence the propagator, is gauge-dependent off-shell; the same factor is what lets the interior cancellation and the on-shell subtraction remove $\Sigma_P$ entirely, independently of the form of $\Omega$.

\subsection{The fermion-photon vertex}

The self-energy is not gauge-invariant on its own, and the Ward--Takahashi identity is the reason: it ties the $(1-\xi)$ content of the self-energy to that of the fermion-photon vertex, so the two must be treated together. We therefore record the one-loop vertex correction here, alongside $\Sigma$. The same longitudinal photon $D^P$ that produces $\Sigma_P$ also dresses the vertex, and \eqref{eq:WTI_tree} fixes the longitudinal projection $(p_2-p_1)_\mu\Gamma^\mu$ in terms of the self-energies $\Sigma_P(p_1)$ and $\Sigma_P(p_2)$ on the two lines that meet at it. Writing $p_1$ for the momentum entering the vertex and $p_2$ for the one leaving, the pinch part of the correction is
\begin{equation}
  \Gamma_P^\mu(p_2,p_1) = g^2 C_2 \dL \gamma^\rho\, \Sf(p_2-\ell)\, \gamma^\mu\, \Sf(p_1-\ell)\, \gamma^\sigma\, D^P_{\rho\sigma}(\ell),
  \label{eq:vertex_def}
\end{equation}
where the subscript $P$ marks the pinch part and distinguishes it from the full proper vertex $\Gamma^\mu(k',k)$ later. 

\subsection{The pinch technique}

Considered in isolation, the self-energy has no reason to be gauge-invariant, because it is not a physical observable. But it always appears as an internal line in a physical amplitude. Consider the simplest such embedding, a fermion propagator $\Sf(q)$ dressed by $\Sigma(q)$, with external fermion propagators $\Sf(p')$ and $\Sf(p)$ attached on either side. The gauge-dependent part of the dressed amplitude takes the schematic form
\begin{equation}
  \Sf(p') \cdots \Sigma_P(q) \cdots \Sf(p),
\end{equation}
where the dots represent the photon vertices and other Dirac structures of the process.

The $\ell^\mu$ from the pinch photon $D^P_{\mu\nu}$ in (\ref{eq:DFandDP}) now inserts a $\slashed\ell$ on the fermion line at each endpoint of the internal photon. When $\slashed\ell$ is inserted between two adjacent fermion propagators, the identity
\begin{equation}
  \Sf(k) \slashed\ell \Sf(k+\ell)
  = i\bigl[\Sf(k)-\Sf(k+\ell)\bigr]
  \label{eq:fund_intro}
\end{equation}
fires, and $\slashed\ell$ splits into two inverse propagators, each of which eats its neighbor, leaving a {\em difference of propagators}.

In the Feynman diagram, the cancellation $\SfI \Sf = \mathbf{1}$ amounts to ``pinching off'' one of the adjacent fermion propagators (hence the name pinch technique). For each contracted $\slashed\ell$ the amplitude splits according to which side of the vertex was pinched off,
\begin{equation}
 {\cal M}_P \to {\cal M}_L + {\cal M}_R.
\end{equation}

The two pieces carry definite signs. Reading \eqref{eq:fund_intro}, the $+i\Sf(k)$ term comes from the $+i\SfI(k+\ell)$ piece of the WTI $\slashed\ell = i[\SfI(k+\ell)-\SfI(k)]$ eating the adjacent $\Sf(k+\ell)$ on its right, so the right propagator $\Sf(k+\ell)$ is pinched off and the left $\Sf(k)$ survives; the $-i\Sf(k+\ell)$ term comes from the $-i\SfI(k)$ piece eating the $\Sf(k)$ on its left, so the left propagator is pinched off and the right survives. The sign rule is therefore
\begin{equation}
  \text{pinch off the right propagator} \to +,\qquad
  \text{pinch off the left propagator} \to -.
  \label{eq:pinch_signs}
\end{equation}
Equivalently, the propagator on the ``outgoing'' side of the photon insertion (where the momentum has been incremented by $+\ell$) is pinched with sign $+$; the ``incoming'' side with sign $-$. At the other endpoint of the same pinch photon the photon momentum reverses, so the geometric assignment of left/right swaps, but the algebraic rule is unchanged.

The identity \eqref{eq:fund_intro} is the tree-level Ward--Takahashi identity in disguise. Its content is local to the segment, in that it does not depend on what lies to the left or right of the two propagators, whether it is another propagator, an on-shell spinor or an arbitrary Dirac structure. The identity fires identically in every context, and it never crosses the external photon vertex that bounds the segment.

\subsection{Boundary terms and the Compton extension}

When the fermion propagator is embedded in a process with external photon vertices, such as Compton scattering, the $\slashed\ell$ insertion can also land on the segments adjacent to the external legs. The WTI still fires locally at each insertion point, but now the telescoping may reach the {\em boundary} of the fermion line. There, $\SfI$ finds no further propagator to cancel against, and a boundary term survives, in which the external-leg propagator is shifted, $\Sf(p) \to \Sf(p-\ell)$, while the rest of the amplitude is untouched.

These boundary terms are gauge-dependent, being proportional to $(1-\xi)$, and they must be cancelled for the full amplitude to be gauge-invariant. The structure of the WTI itself points to what cancels them. A boundary term at the $p$-leg has the form of a self-energy insertion on that leg, with only $\Sf(p)$ modified. Similarly, boundary terms where both legs are shifted correspond to a doubly-shifted tree, which is cancelled by the crossed ($u$-channel) diagram. None of these boundary cancellations affect $\hat\Sigma$, which depends only on the internal momentum $q$; the $u$-channel's role is purely in the boundary sector.

The $s$-channel Compton amplitude, with its two external photon vertices, is the {\em minimal process} that generates all structurally distinct configurations, namely the self-energy topology (no spectator photon), the vertex topology (one spectator photon) and the box topology (two spectator photons). The $u$-channel is needed only to complete the boundary cancellation of the doubly-shifted tree, not to define $\hat\Sigma$. Every higher-point process produces the same cancellation mechanisms with additional spectator photons that the WTI does not touch.

\section{The Off-Shell Compton Amplitude}
\label{sec:explicit}

\subsection{Compton and the general case}

We are mainly interested in the off-shell mass, whose loop correction is the self-energy, which for a fermion reduces to the Compton process. This already suffices to demonstrate the {\em local mechanism}, whereby at each vertex the WTI fires independently within its own segment, producing a difference of propagators that telescopes without crossing the segment boundary. The three structurally distinct local configurations are
\begin{enumerate}
\item {\em Self-energy.} Both insertions on the same segment. Pure internal telescoping, no spectator photon vertices.
\item {\em Vertex.} One insertion internal, one external. Internal telescoping on one side, boundary term on the other. One spectator photon vertex.
\item {\em Box.} Both insertions on external legs. Boundary terms at both ends. Two spectator photon vertices.
\end{enumerate}

However, the local mechanism alone does not complete the cancellation for processes with more than two external photon vertices. Topologies with more external photons between the pinch-photon endpoints arise (pentagons, hexagons, etc.), and after the WTI fires at both endpoints, each topology produces shifted-boundary terms that do not cancel within that single topology. We give the explicit demonstration in Section~\ref{sec:general_n}, after the detailed Compton calculation below.

By power counting in QED, only the 2-point and 3-point functions require renormalization. Pentagon and higher topologies are UV-finite; they do not contribute to the renormalization of the self-energy or vertex but participate in the $k$-telescoping that ensures gauge invariance of the full amplitude.

\subsection{The $s$-channel topologies}

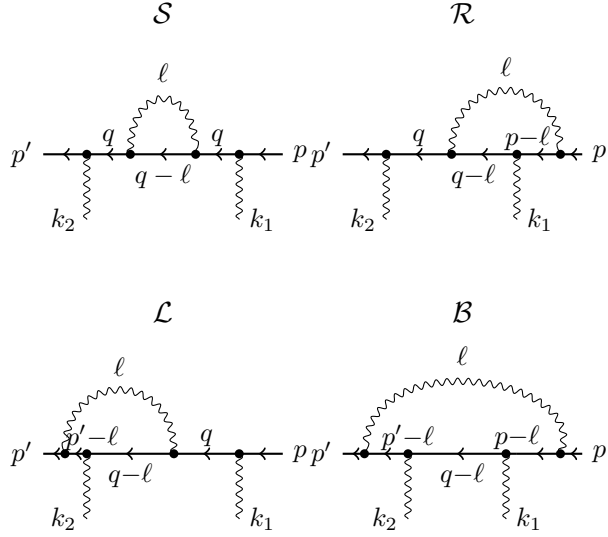
\begin{figure}[t]
\centering
\begin{tikzpicture}[
  fermion/.style={thick, postaction={decorate,
    decoration={markings, mark=at position 0.55 with {\arrow{>}}}}},
  photon/.style={decorate, decoration={snake, amplitude=1.2pt,
    segment length=4pt}},
  pinchedphoton/.style={densely dotted, thick},
  vertex/.style={fill, circle, inner sep=1.2pt},
  scale=0.72
]
\begin{scope}[shift={(0,0)}]
  \node at (2.2,2.6) {$\mathcal{S}$};
  \draw[fermion] (4.4,0) node[right]{\small$p$} -- (3.6,0);
  \draw[fermion] (3.6,0) -- (2.8,0);
  \draw[fermion] (2.8,0) -- (1.6,0);
  \draw[fermion] (1.6,0) -- (0.8,0);
  \draw[fermion] (0.8,0) -- (0,0) node[left]{\small$p'$};
  \node[vertex] at (3.6,0) {};
  \node[vertex] at (0.8,0) {};
  \node[vertex] at (1.6,0) {};
  \node[vertex] at (2.8,0) {};
  \draw[photon] (3.6,0) -- (3.6,-1.2) node[right]{\small$k_1$};
  \draw[photon] (0.8,0) -- (0.8,-1.2) node[left]{\small$k_2$};
  \draw[photon] (1.6,0) .. controls (1.6,1.3) and (2.8,1.3) .. (2.8,0);
  \node at (2.2,1.5) {\small$\ell$};
  \node at (2.2,-0.4) {\small$q-\ell$};
  \node at (3.2,0.3) {\small$q$};
  \node at (1.2,0.3) {\small$q$};
\end{scope}

\begin{scope}[shift={(5.5,0)}]
  \node at (2.2,2.6) {$\mathcal{R}$};
  \draw[fermion] (4.4,0) node[right]{\small$p$} -- (4.0,0);
  \draw[fermion] (4.0,0) -- (3.2,0);
  \draw[fermion] (3.2,0) -- (2.0,0);
  \draw[fermion] (2.0,0) -- (0.8,0);
  \draw[fermion] (0.8,0) -- (0,0) node[left]{\small$p'$};
  \node[vertex] at (3.2,0) {};
  \node[vertex] at (0.8,0) {};
  \node[vertex] at (2.0,0) {};
  \node[vertex] at (4.0,0) {};
  \draw[photon] (3.2,0) -- (3.2,-1.2) node[right]{\small$k_1$};
  \draw[photon] (0.8,0) -- (0.8,-1.2) node[left]{\small$k_2$};
  \draw[photon] (2.0,0) .. controls (2.0,1.5) and (4.0,1.5) .. (4.0,0);
  \node at (3.0,1.6) {\small$\ell$};
  \node at (3.4,0.3) {\small$p{-}\ell$};
  \node at (2.4,-0.4) {\small$q{-}\ell$};
  \node at (1.4,0.3) {\small$q$};
\end{scope}

\begin{scope}[shift={(0,-5.5)}]
  \node at (2.2,2.6) {$\mathcal{L}$};
  \draw[fermion] (4.4,0) node[right]{\small$p$} -- (3.6,0);
  \draw[fermion] (3.6,0) -- (2.4,0);
  \draw[fermion] (2.4,0) -- (0.8,0);
  \draw[fermion] (0.8,0) -- (0.4,0);
  \draw[fermion] (0.4,0) -- (0,0) node[left]{\small$p'$};
  \node[vertex] at (3.6,0) {};
  \node[vertex] at (0.8,0) {};
  \node[vertex] at (2.4,0) {};
  \node[vertex] at (0.4,0) {};
  \draw[photon] (3.6,0) -- (3.6,-1.2) node[right]{\small$k_1$};
  \draw[photon] (0.8,0) -- (0.8,-1.2) node[left]{\small$k_2$};
  \draw[photon] (2.4,0) .. controls (2.4,1.5) and (0.4,1.5) .. (0.4,0);
  \node at (1.4,1.6) {\small$\ell$};
  \node at (3.0,0.3) {\small$q$};
  \node at (1.6,-0.4) {\small$q{-}\ell$};
  \node at (0.9,0.3) {\small$p'{-}\ell$};
\end{scope}

\begin{scope}[shift={(5.5,-5.5)}]
  \node at (2.2,2.6) {$\mathcal{B}$};
  \draw[fermion] (4.4,0) node[right]{\small$p$} -- (4.0,0);
  \draw[fermion] (4.0,0) -- (3.2,0);
  \draw[fermion] (3.2,0) -- (1.2,0);
  \draw[fermion] (1.2,0) -- (0.4,0);
  \draw[fermion] (0.4,0) -- (0,0) node[left]{\small$p'$};
  \node[vertex] at (3,0) {};
  \node[vertex] at (1.2,0) {};
  \node[vertex] at (4.0,0) {};
  \node[vertex] at (0.4,0) {};
  \draw[photon] (3,0) -- (3,-1.2) node[right]{\small$k_1$};
  \draw[photon] (1.2,0) -- (1.2,-1.2) node[left]{\small$k_2$};
  \draw[photon] (4.0,0) .. controls (4.5,1.4) and (0.4,2.0) .. (0.4,0);
  \node at (2.2,1.8) {\small$\ell$};
  \node at (3.2,0.3) {\small$p{-}\ell$};
  \node at (2.2,-0.4) {\small$q{-}\ell$};
  \node at (1.2,0.3) {\small$p'{-}\ell$};
\end{scope}

\end{tikzpicture}
\caption{The four $s$-channel topologies of the one-loop corrected Compton process before pinching. They are classified by which segments of the fermion line carry the two endpoints of the virtual photon $\ell$, namely the intermediate-line self-energy $\mathcal{S}$, the right and left vertex corrections $\mathcal{R}$ and $\mathcal{L}$ and the box $\mathcal{B}$.}
\label{fig:topologies}
\end{figure}

The $s$-channel Compton amplitude at tree level has the fermion-line structure 
\begin{equation}
\Sf(p') \slashed\epsilon_2^* \Sf(q) \slashed\epsilon_1 \Sf(p), \qquad q = p+k_1 = p'-k_2.
\end{equation}
We call the fermion propagator between the two photon vertices, carrying momentum $q$ at tree level, the {\em intermediate fermion line}. At one loop, a virtual photon with momentum $\ell$ connects two points on the fermion line. There are four topologies, distinguished by which segments the virtual photon endpoints sit on, namely the intermediate-line self-energy $\mathcal{S}$, the vertex corrections $\mathcal{R}$ and $\mathcal{L}$ and the box $\mathcal{B}$ (shown in Fig.~\ref{fig:topologies}). We may also consider leg self-energies in the legs.

\begin{figure}[ht]
\centering
\begin{tikzpicture}[
  fermion/.style={thick, postaction={decorate,
    decoration={markings, mark=at position 0.55 with {\arrow{>}}}}},
  photon/.style={decorate, decoration={snake, amplitude=0.8pt,
    segment length=3pt}},
  pinchedphoton/.style={densely dotted, thick},
  vertex/.style={fill, circle, inner sep=0.9pt},
  scale=0.85,
  every node/.style={font=\scriptsize}
]

\begin{scope}[shift={(0,0)}]
  \node at (1.1,1.4) {$\mathcal{S}_{LL}=+(2,2)$};
  \draw[fermion] (2.2,0) node[right]{$p$} -- (0,0) node[left]{$p'$};
  \node[vertex] at (0.5,0){}; \node[vertex] at (1.7,0){};
  \draw[photon] (0.5,0) -- (0.5,-0.5); \draw[photon] (1.7,0) -- (1.7,-0.5);
  \draw[pinchedphoton] (0.5,0) .. controls (0.15,0.9) and (0.85,0.9) .. (0.5,0);
  \node[font=\tiny] at (1.1,-0.26) {$q$};
\end{scope}
\begin{scope}[shift={(3.6,0)}]
  \node at (1.1,1.4) {$\mathcal{S}_{LR}=-(1,2)$};
  \draw[fermion] (2.2,0) node[right]{$p$} -- (0,0) node[left]{$p'$};
  \node[vertex] at (0.5,0){}; \node[vertex] at (1.7,0){};
  \draw[photon] (0.5,0) -- (0.5,-0.5); \draw[photon] (1.7,0) -- (1.7,-0.5);
  \draw[pinchedphoton] (0.5,0) .. controls (0.7,0.95) and (1.5,0.95) .. (1.7,0);
  \node[font=\tiny] at (1.1,-0.26) {$q-\ell$};
\end{scope}
\begin{scope}[shift={(7.2,0)}]
  \node at (1.1,1.4) {$\mathcal{S}_{RL}=-(2,2)$};
  \draw[fermion] (2.2,0) node[right]{$p$} -- (0,0) node[left]{$p'$};
  \node[vertex] at (0.5,0){}; \node[vertex] at (1.7,0){};
  \draw[photon] (0.5,0) -- (0.5,-0.5); \draw[photon] (1.7,0) -- (1.7,-0.5);
  \draw[pinchedphoton] (0.5,0) .. controls (0.15,0.9) and (0.85,0.9) .. (0.5,0);
  \node[font=\tiny] at (1.1,-0.26) {$q$};
\end{scope}
\begin{scope}[shift={(10.8,0)}]
  \node at (1.1,1.4) {$\mathcal{S}_{RR}=+(1,1)$};
  \draw[fermion] (2.2,0) node[right]{$p$} -- (0,0) node[left]{$p'$};
  \node[vertex] at (0.5,0){}; \node[vertex] at (1.7,0){};
  \draw[photon] (0.5,0) -- (0.5,-0.5); \draw[photon] (1.7,0) -- (1.7,-0.5);
  \draw[pinchedphoton] (1.7,0) .. controls (1.35,0.9) and (2.05,0.9) .. (1.7,0);
  \node[font=\tiny] at (1.1,-0.26) {$q$};
\end{scope}

\begin{scope}[shift={(0,-2.6)}]
  \node at (1.1,1.4) {$\mathcal{R}_{LL}=+(1,2)$};
  \draw[fermion] (2.2,0) node[right]{$p$} -- (0,0) node[left]{$p'$};
  \node[vertex] at (0.5,0){}; \node[vertex] at (1.7,0){};
  \draw[photon] (0.5,0) -- (0.5,-0.5); \draw[photon] (1.7,0) -- (1.7,-0.5);
  \draw[pinchedphoton] (0.5,0) .. controls (0.7,0.95) and (1.5,0.95) .. (1.7,0);
  \node[font=\tiny] at (1.1,-0.26) {$q-\ell$};
  \node[font=\tiny] at (1.95,-0.26) {$p$};
\end{scope}
\begin{scope}[shift={(3.6,-2.6)}]
  \node at (1.1,1.4) {$\mathcal{R}_{LR}=-(+,2)$};
  \draw[fermion] (2.2,0) node[right]{$p$} -- (0,0) node[left]{$p'$};
  \node[vertex] at (0.5,0){}; \node[vertex] at (1.7,0){};
  \draw[photon] (0.5,0) -- (0.5,-0.5); \draw[photon] (1.7,0) -- (1.7,-0.5);
  \draw[pinchedphoton] (0.5,0) .. controls (0.9,1.05) and (1.9,1.05) .. (2.2,0);
  \node[font=\tiny] at (1.1,-0.26) {$q-\ell$};
  \node[font=\tiny] at (1.95,-0.26) {$p-\ell$};
\end{scope}
\begin{scope}[shift={(7.2,-2.6)}]
  \node at (1.1,1.4) {$\mathcal{R}_{RL}=-(1,1)$};
  \draw[fermion] (2.2,0) node[right]{$p$} -- (0,0) node[left]{$p'$};
  \node[vertex] at (0.5,0){}; \node[vertex] at (1.7,0){};
  \draw[photon] (0.5,0) -- (0.5,-0.5); \draw[photon] (1.7,0) -- (1.7,-0.5);
  \draw[pinchedphoton] (1.7,0) .. controls (1.35,0.9) and (2.05,0.9) .. (1.7,0);
  \node[font=\tiny] at (1.1,-0.26) {$q$};
  \node[font=\tiny] at (1.95,-0.26) {$p$};
\end{scope}
\begin{scope}[shift={(10.8,-2.6)}]
  \node at (1.1,1.4) {$\mathcal{R}_{RR}=+(+,1)$};
  \draw[fermion] (2.2,0) node[right]{$p$} -- (0,0) node[left]{$p'$};
  \node[vertex] at (0.5,0){}; \node[vertex] at (1.7,0){};
  \draw[photon] (0.5,0) -- (0.5,-0.5); \draw[photon] (1.7,0) -- (1.7,-0.5);
  \draw[pinchedphoton] (1.7,0) .. controls (1.85,0.85) and (2.1,0.6) .. (2.2,0);
  \node[font=\tiny] at (1.1,-0.26) {$q$};
  \node[font=\tiny] at (1.95,-0.26) {$p-\ell$};
\end{scope}

\begin{scope}[shift={(0,-5.2)}]
  \node at (1.1,1.4) {$\mathcal{L}_{LL}=+(2,-)$};
  \draw[fermion] (2.2,0) node[right]{$p$} -- (0,0) node[left]{$p'$};
  \node[vertex] at (0.5,0){}; \node[vertex] at (1.7,0){};
  \draw[photon] (0.5,0) -- (0.5,-0.5); \draw[photon] (1.7,0) -- (1.7,-0.5);
  \draw[pinchedphoton] (0,0) .. controls (0.1,0.6) and (0.35,0.85) .. (0.5,0);
  \node[font=\tiny] at (0.32,-0.28) {$p'-\ell$};
  \node[font=\tiny] at (1.1,-0.26) {$q$};
\end{scope}
\begin{scope}[shift={(3.6,-5.2)}]
  \node at (1.1,1.4) {$\mathcal{L}_{LR}=-(1,-)$};
  \draw[fermion] (2.2,0) node[right]{$p$} -- (0,0) node[left]{$p'$};
  \node[vertex] at (0.5,0){}; \node[vertex] at (1.7,0){};
  \draw[photon] (0.5,0) -- (0.5,-0.5); \draw[photon] (1.7,0) -- (1.7,-0.5);
  \draw[pinchedphoton] (0,0) .. controls (0.3,1.05) and (1.3,1.05) .. (1.7,0);
  \node[font=\tiny] at (0.32,-0.28) {$p'-\ell$};
  \node[font=\tiny] at (1.1,-0.26) {$q-\ell$};
\end{scope}
\begin{scope}[shift={(7.2,-5.2)}]
  \node at (1.1,1.4) {$\mathcal{L}_{RL}=-(2,2)$};
  \draw[fermion] (2.2,0) node[right]{$p$} -- (0,0) node[left]{$p'$};
  \node[vertex] at (0.5,0){}; \node[vertex] at (1.7,0){};
  \draw[photon] (0.5,0) -- (0.5,-0.5); \draw[photon] (1.7,0) -- (1.7,-0.5);
  \draw[pinchedphoton] (0.5,0) .. controls (0.15,0.9) and (0.85,0.9) .. (0.5,0);
  \node[font=\tiny] at (0.25,-0.26) {$p'$};
  \node[font=\tiny] at (1.1,-0.26) {$q$};
\end{scope}
\begin{scope}[shift={(10.8,-5.2)}]
  \node at (1.1,1.4) {$\mathcal{L}_{RR}=+(1,2)$};
  \draw[fermion] (2.2,0) node[right]{$p$} -- (0,0) node[left]{$p'$};
  \node[vertex] at (0.5,0){}; \node[vertex] at (1.7,0){};
  \draw[photon] (0.5,0) -- (0.5,-0.5); \draw[photon] (1.7,0) -- (1.7,-0.5);
  \draw[pinchedphoton] (0.5,0) .. controls (0.7,0.95) and (1.5,0.95) .. (1.7,0);
  \node[font=\tiny] at (1.1,-0.26) {$q-\ell$};
\end{scope}

\begin{scope}[shift={(0,-7.8)}]
  \node at (1.1,1.4) {$\mathcal{B}_{LL}=+(1,-)$};
  \draw[fermion] (2.2,0) node[right]{$p$} -- (0,0) node[left]{$p'$};
  \node[vertex] at (0.5,0){}; \node[vertex] at (1.7,0){};
  \draw[photon] (0.5,0) -- (0.5,-0.5); \draw[photon] (1.7,0) -- (1.7,-0.5);
  \draw[pinchedphoton] (0,0) .. controls (0.3,1.05) and (1.3,1.05) .. (1.7,0);
  \node[font=\tiny] at (0.32,-0.28) {$p'-\ell$};
  \node[font=\tiny] at (1.1,-0.26) {$q-\ell$};
  \node[font=\tiny] at (1.95,-0.26) {$p$};
\end{scope}
\begin{scope}[shift={(3.6,-7.8)}]
  \node at (1.1,1.4) {$\mathcal{B}_{LR}=-(+,-)$};
  \draw[fermion] (2.2,0) node[right]{$p$} -- (0,0) node[left]{$p'$};
  \node[vertex] at (0.5,0){}; \node[vertex] at (1.7,0){};
  \draw[photon] (0.5,0) -- (0.5,-0.5); \draw[photon] (1.7,0) -- (1.7,-0.5);
  \draw[pinchedphoton] (0,0) .. controls (0.4,1.25) and (1.8,1.25) .. (2.2,0);
  \node[font=\tiny] at (0.32,-0.28) {$p'-\ell$};
  \node[font=\tiny] at (1.1,-0.26) {$q-\ell$};
  \node[font=\tiny] at (1.95,-0.26) {$p-\ell$};
\end{scope}
\begin{scope}[shift={(7.2,-7.8)}]
  \node at (1.1,1.4) {$\mathcal{B}_{RL}=-(1,2)$};
  \draw[fermion] (2.2,0) node[right]{$p$} -- (0,0) node[left]{$p'$};
  \node[vertex] at (0.5,0){}; \node[vertex] at (1.7,0){};
  \draw[photon] (0.5,0) -- (0.5,-0.5); \draw[photon] (1.7,0) -- (1.7,-0.5);
  \draw[pinchedphoton] (0.5,0) .. controls (0.7,0.95) and (1.5,0.95) .. (1.7,0);
  \node[font=\tiny] at (1.1,-0.26) {$q-\ell$};
\end{scope}
\begin{scope}[shift={(10.8,-7.8)}]
  \node at (1.1,1.4) {$\mathcal{B}_{RR}=+(+,2)$};
  \draw[fermion] (2.2,0) node[right]{$p$} -- (0,0) node[left]{$p'$};
  \node[vertex] at (0.5,0){}; \node[vertex] at (1.7,0){};
  \draw[photon] (0.5,0) -- (0.5,-0.5); \draw[photon] (1.7,0) -- (1.7,-0.5);
  \draw[pinchedphoton] (0.5,0) .. controls (0.9,1.05) and (1.9,1.05) .. (2.2,0);
  \node[font=\tiny] at (1.1,-0.26) {$q-\ell$};
  \node[font=\tiny] at (1.95,-0.26) {$p-\ell$};
\end{scope}
\end{tikzpicture}
\caption{Pinched diagrams for the four topologies. At each $D^P$ vertex the WTI removes (pinches off) the fermion propagator on one side, so that vertex collapses onto its neighbor; the dotted line is the resulting longitudinal photon, drawn between the two collapse points, and the surviving propagator momenta are labeled. The two independent left/right choices at the two vertices generate four terms per topology, labeled by the pinch direction at the left and right vertices, namely $LL$, $LR$, $RL$ and $RR$. These are the interior sector ($RL$), the doubly-shifted sector ($LR$) and the two single-boundary shifts ($LL$ and $RR$).}
\label{fig:pinched}
\end{figure}
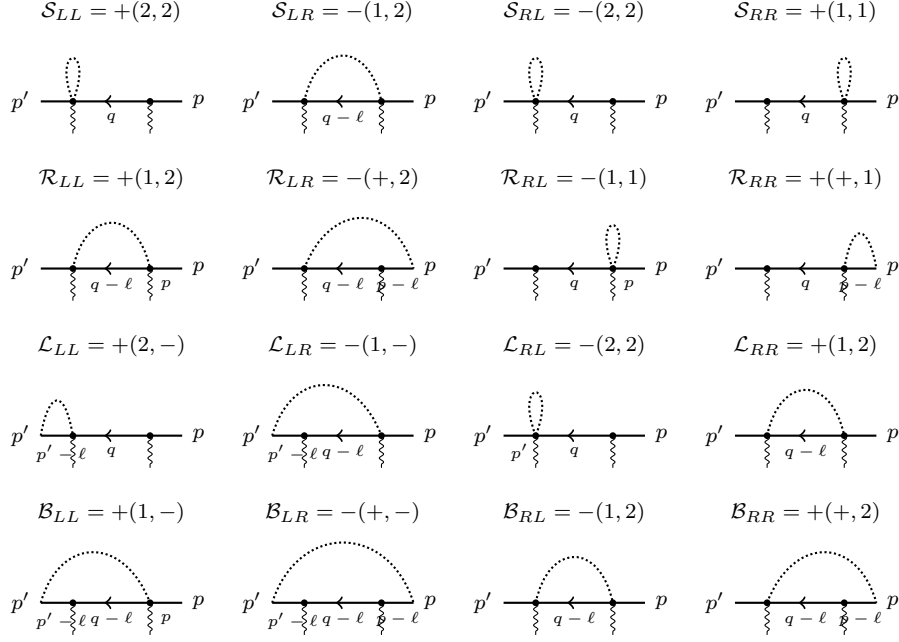

The full one-loop amplitudes, retaining only the pinch part $D^P_{\mu\nu}(\ell)$ of the virtual photon propagator in (\ref{eq:DFandDP}), are the following.

\paragraph{Self-energy on the intermediate line $\mathcal{S}$} It has both virtual-photon vertices on the internal segment between $\epsilon_1$ and $\epsilon_2$
\begin{equation}
\begin{split}
 i \mathcal{M}_{\mathcal{S}}^P
  &= e^2g^2 C_2\dL \Sf(p') \slashed\epsilon_2^*
    \Sf(q)\gamma^\mu \Sf(q-\ell) \gamma^\nu \Sf(q)
     \slashed\epsilon_1 \Sf(p)
    D^P_{\mu\nu}(\ell) \\
    & = e^2 \Sf(p') \slashed\epsilon_2^* \Sf(q) \Sigma_P(q) \Sf(q)
     \slashed\epsilon_1 \Sf(p) .
\end{split}
  \label{eq:MS} 
\end{equation}

\paragraph{Right vertex correction $\mathcal{R}$} It has one virtual-photon vertex on the $q$-segment and the other on the $p$-leg
\begin{equation}
\begin{split}
  i \mathcal{M}_{\mathcal{R}}^P
  &= e^2g^2 C_2\dL \Sf(p') \slashed\epsilon_2^* 
    \Sf(q) \gamma^\mu \Sf(q-\ell) \slashed\epsilon_1 
    \Sf(p-\ell) \gamma^\nu \Sf(p)
    D^P_{\mu\nu}(\ell) \\
    & = e^2 \Sf(p') \slashed\epsilon_2^* 
    \Sf(q) \Gamma_P^\nu \epsilon_{1,\nu} \Sf(p),
  \end{split}
  \label{eq:MR}
\end{equation}
where the second line uses the right vertex $\Gamma_P^\nu(q,p)$ of \eqref{eq:vertex_def}.

\paragraph{Left vertex correction $\mathcal{L}$} It has one virtual-photon vertex on the $q$-segment and the other on the $p'$-leg
\begin{equation}
\begin{split}
  i \mathcal{M}_{\mathcal{L}}^P
  &= e^2g^2 C_2\dL \Sf(p') \gamma^\mu \Sf(p'-\ell) 
    \slashed\epsilon_2^* \Sf(q-\ell) \gamma^\nu 
    \Sf(q) \slashed\epsilon_1 \Sf(p)
    D^P_{\mu\nu}(\ell) \\
   & = e^2 \Sf(p') \bar\Gamma_P^\mu \epsilon_{2,\mu}^*
    \Sf(q) \slashed\epsilon_1 \Sf(p),
  \end{split}
  \label{eq:ML}
\end{equation}
where the second line uses the left vertex $\bar\Gamma_P^\mu(p',q)\equiv\Gamma_P^\mu(p',q)$ of \eqref{eq:vertex_def}.

\paragraph{Box $\mathcal{B}$} It has one virtual-photon vertex on the $p$-leg and the other on the $p'$-leg
\begin{equation}
 i  \mathcal{M}_{\mathcal{B}}^P
  = e^2g^2 C_2\dL \Sf(p') \gamma^\mu \Sf(p'-\ell) 
    \slashed\epsilon_2^* \Sf(q-\ell) 
    \slashed\epsilon_1 \Sf(p-\ell) \gamma^\nu \Sf(p)
    D^P_{\mu\nu}(\ell).
  \label{eq:MB}
\end{equation}

In each amplitude, the $\ell^\mu$ and $\ell^\nu$ from $D^P_{\mu\nu}$ contract with the adjacent $\gamma^\mu$ and $\gamma^\nu$, replacing them by $\slashed\ell$ insertions. The fundamental identity \eqref{eq:fund_intro} then fires at each insertion point, producing the pinched amplitudes below.

\subsection{Segments and segment locality}
\label{sec:segments}

Consider a fermion line carrying a sequence of propagators separated by $n+1$ external photon vertices $\slashed\epsilon_{a_i}$
\begin{equation}
  \Sf(p') \slashed\epsilon_{a_{n+1}} \cdots
   \slashed\epsilon_{a_2} \Sf(q_1) \slashed\epsilon_{a_1} \Sf(p).
\end{equation}
The external photon vertices partition the fermion line into $n+2$ {\em segments}, namely the $q_0$-segment (from the right ($p$) fermion endpoint to $\epsilon_{a_1}$), the $q_1$-segment (from $\epsilon_{a_1}$ to $\epsilon_{a_2}$) and so on, segment $s$ carrying momentum $q_s$ with the outermost legs $q_0 = p$ and $q_{n+1} = p'$. Each segment contains exactly one propagator $\Sf(q_s)$ (or, at one loop, a string of propagators connected by the pinch-photon vertex).

In general, when a gauge boson line forms a loop by ending at a pair of segments $[s_1, s_2]$ with $0 \le s_1 \le s_2 \le n+1$, the corresponding amplitude involving the longitudinal part $D^P$ becomes
\begin{equation}
\begin{split}
 i \mathcal{M}^P_{[s_1, s_2]}
  &= (-i e)^{n+1} (-i g)^2 C_2 \dL\;
    \Sf(p')\cdots
    \Sf(q_{s_2})\,\gamma^\nu\,\Sf(q_{s_2}+\ell) \\
 & \qquad \qquad  \qquad \cdots\slashed\epsilon_{a_{s_1+1}}\,
    \Sf(q_{s_1}+\ell)\,\gamma^\mu\,\Sf(q_{s_1})
    \cdots\slashed\epsilon_{a_1}\Sf(p)\;
    D^P_{\mu\nu}(\ell),
    \end{split}
  \label{eq:placement_amp}
\end{equation}
with $D^P_{\mu\nu}(\ell)$ in (\ref{eq:DFandDP}) and $\gamma^\mu, \gamma^\nu$ the two pinch-photon vertices on segments $s_1$ and $s_2$; the propagators between them carry the shift $\ell$. The self-energy insertion $\Sf(q_s)\gamma^\nu\Sf(q_s+\ell)\gamma^\mu\Sf(q_s)$ corresponds to $s = s_1 = s_2$.
Then the above four Compton topologies are the placements
\begin{equation}
  \mathcal{R} = [0,1], \ 
  \mathcal{B} = [0,2], \ 
  \mathcal{S} = [1,1], \ 
  \mathcal{L} = [1,2].
  \label{eq:topology_placements}
\end{equation}
In what follows we use $\mathcal{M}_X^P$ and $X={\cal R, B, S, L}$ interchangeably.

\subsection{Pinching amplitudes}

Inserting the pinch part $D^P_{\mu\nu}$ of \eqref{eq:DFandDP} into the amplitude contracts $\ell^\mu$ and $\ell^\nu$ with the two pinch-photon vertices, producing one $\slashed\ell$ at each. The fundamental identity \eqref{eq:fund_intro} then fires at each insertion
\begin{align}
  \text{Left insertion: }&
  \Sf(q_i) \slashed\ell \Sf(q_i+\ell)
  = i\bigl[\Sf(q_i)-\Sf(q_i+\ell)\bigr],
  \label{eq:left_ins}\\[4pt]
  \text{Right insertion: }&
  \Sf(q_j+\ell) \slashed\ell \Sf(q_j)
  = i\bigl[\Sf(q_j+\ell)-\Sf(q_j)\bigr].
  \label{eq:right_ins}
\end{align}
The result is a difference of propagators, both of which carry momenta belonging to the \textit{same} segment. The external photon vertex $\slashed\epsilon_{a_i}$ at the segment boundary is a spectator, since the identity does not touch it and the telescoping cannot cross it.

The two firings together split the amplitude into four sectors,
\begin{equation}
  {\cal M}^P = {\cal M}_{LL} + {\cal M}_{LR} + {\cal M}_{RL} + {\cal M}_{RR},
  \label{eq:four_sectors}
\end{equation}
where each superscript records, at each of the two $\slashed\ell$ insertions, which bounding vertex of its segment the surviving endpoint lands on, with ``L'' leaving it at the higher-index vertex $\gamma_{s+1}$ (nearer $p'$) and ``R'' at the lower-index $\gamma_s$ (nearer $p$). Diagrammatically, each choice removes one of the adjacent internal fermion lines, and the four products are the four shift outcomes. Geometrically, $RL$ is the interior outcome, both pinches going inward (toward the middle of the block) to remove the interior propagators and leave the boundaries untouched; $LR$ is the doubly-shifted outcome, both pinches going outward to catch both boundaries; $LL$ and $RR$ are the two single-boundary shifts, mixed firings that each catch one boundary. This is the {\em locality of the segment}, meaning that when the two pinch endpoints lie on different segments, each fires independently within its own segment, and no term involves a propagator shift that crosses a segment boundary. It is this locality that lets the gauge-dependent part of the self-energy on each fermion interval be removed by the firings at the two vertices bounding that interval and by nothing else, so that the self-energy is gauge-invariant interval by interval, independently of the rest of the diagram.

The shifted propagators are those along the fermion-line path between $a$ and $b$, so all four sectors at the same placement have distinct $(a,b)$ and distinct Dirac structures. Explicitly, for $a \le b$ the endpoint pair contributes the full pinch amplitude $\mathcal{M}^P_{(a,b)}$, the tree chain carrying the shift $\ell$ on segments $a, \ldots, b-1$ dressed with the loop measure and the residual factor $1/\ell^4$ that the two firings leave behind,
\begin{equation}
\begin{split}
 i \mathcal{M}^P_{(a,b)}
  &= (1-\xi) (-i e)^{n+1} (-ig)^2 C_2 \dL\,\frac{1}{\ell^4}     \Sf(p')\cdots\Sf(q_b)\,\slashed\epsilon_{a_b}    \Sf(q_{b-1}+\ell)\,\\
    & \qquad \qquad \cdots\slashed\epsilon_{a_{a+1}}   \Sf(q_a+\ell)\,\slashed\epsilon_{a_a} \Sf(q_{a-1})\cdots\slashed\epsilon_{a_1}\Sf(p),
  \label{eq:ab_amp}
\end{split}
\end{equation}
all other propagators unshifted, with the endpoint conventions $\gamma_0 = +$ (the $p$ endpoint) and $\gamma_{n+2} = -$ (the $p'$ endpoint). The prefactor and $1/\ell^4$ are common to every endpoint pair and match the one-loop forms~\eqref{eq:DeltaGammaR} and~\eqref{eq:DeltaGammaL} given below. We abbreviate $\mathcal{M}^P_{(a,b)}$ by $(a,b)$, attaching a sector label $(a,b)_{LL}$, $(a,b)_{LR}$, $(a,b)_{RL}$ or $(a,b)_{RR}$ when the firing sector is needed. In particular, when $a=b$, no propagator depends on the internal momentum $\ell$
\begin{equation}
\begin{split}
 i \mathcal{M}^P_{(a,a)} 
  &=- (1-\xi) (-i e)^{n+1} (-ig)^2 C_2  \Sf(p') \slashed\epsilon_{a_{n+1}} \cdots\slashed\epsilon_{a_1}\Sf(p) \dL\,\frac{1}{\ell^4}  ,
  \label{indep_amp}
\end{split}
\end{equation}
so that the amplitude is the same for any $a$.

Applying the single-firing rule~\eqref{eq:pinch_signs} once at each of the two insertions assigns every sector the product of its two signs, so the four sectors of~\eqref{eq:four_sectors} carry the universal pattern $(+,-,-,+)$ in the order $(LL,LR,RL,RR)$.

\subsection{Post-pinch endpoint pairs}
\label{sec:amplitudes}

The two entries of a pair $(a,b)$ are read directly off the diagram as its two collapse points (Fig.~\ref{fig:pinched}).

Terms with the same endpoint pair $(a,b)$ share the same Dirac structure (the pinch endpoints determine which propagators are shifted) and therefore enter the cancellation analysis together. The four boundary-shift outcomes are an interior-shifted Dirac structure when both $a,b$ are at external photons, a single-boundary shift for $(a,\pm)$ or $(\pm,b)$ and the doubly-shifted tree for $(+,-)$.

The four pinched amplitudes read
\begin{align}
  \mathcal{S} &= +(2,2) - (1,2) - (2,2) + (1,1),
    \label{eq:KS}\\
  \mathcal{R} &= +(1,2) - (+,2) - (1,1) + (+,1),
    \label{eq:KR}\\
  \mathcal{L} &= +(2,-) - (1,-) - (2,2) + (1,2),
    \label{eq:KL}\\
  \mathcal{B} &= +(1,-) - (+,-) - (1,2) + (+,2).
    \label{eq:KB}
\end{align}
The signs are the universal $(+,-,-,+)$ pattern in the $(LL,LR,RL,RR)$ order from Section~\ref{sec:segments}. One entry requires a convention: the $RL$ sector of $\mathcal{S}$ is written as $-(2,2)$ rather than the raw map value $-(2,1)$ of \eqref{eq:placement_to_ab}, because both firings consume the single middle propagator; this trapped-pinch configuration is evaluated below and its residue placed at the $\gamma_2$ endpoint.

The two $(2,2)$ entries in $\mathcal{S}$ are the loop at $\epsilon_2$ ($\mathcal{S}_{LL}$) and the trapped pinch ($\mathcal{S}_{RL}$), both having the same Dirac structure. The trapped configuration arises when both WTI firings consume the single middle propagator $\Sf(q-\ell)$; the residue $-\Sf(q)\SfI(q-\ell)\Sf(q) = -\Sf(q) - i\Sf(q)\slashed\ell\Sf(q)$ has its $\slashed\ell$ piece vanish under $\int d^d\ell/\ell^4$, leaving $-\Sf(q)$, and the two $(2,2)$ terms cancel each other (Fig.~\ref{fig:pinched}). As also noted in (\ref{indep_amp}), the amplitude with the pinched points contracted to a point is the same $(1,1)=(2,2)$.

\subsection{Cancellation organized by Dirac structure}
\label{sec:cancel}

We now collect all the terms from the four pinched amplitudes $\mathcal{S}$, $\mathcal{R}$, $\mathcal{L}$, $\mathcal{B}$ and group them by Dirac structure parameterized by $(a,b)$.

\subsubsection{Interior cancellations}

Four Dirac structures cancel exactly, namely those with total coefficient~$0$. In the endpoint notation of Section~\ref{sec:segments}, each Dirac structure corresponds to a distinct endpoint pair $(a,b)$, and the cancellation collects every term carrying that pair across all four topologies.

\paragraph{The $(1,2)$ pair.}
\begin{equation}
-(1,2)_\mathcal{S}+(1,2)_\mathcal{R}+(1,2)_\mathcal{L}-(1,2)_\mathcal{B} = 0.
  \label{eq:cancel_010}
\end{equation}
This is the central cancellation, where each of the four topologies ($\mathcal{S}$, $\mathcal{R}$, $\mathcal{L}$, $\mathcal{B}$) contributes one diagram in which the pinch photon runs from $\epsilon_2$ to $\epsilon_1$, all four share the Dirac structure ``intermediate propagator shifted, external legs untouched,'' and their coefficients sum to zero.

\paragraph{The $(+,2)$ pair.}
\begin{equation}
  -(+,2)_\mathcal{R}+(+,2)_\mathcal{B} = 0.
  \label{eq:cancel_011}
\end{equation}

\paragraph{The $(1,-)$ pair.}
\begin{equation}
 -(1,-)_\mathcal{L} +(1,-)_\mathcal{B} = 0.
  \label{eq:cancel_110}
\end{equation}

\paragraph{The $(1,1)$ pair.}
\begin{equation}
 +(1,1)_\mathcal{S} -(1,1)_\mathcal{R} = 0,
  \label{eq:cancel_diag}
\end{equation}
the diagonal instance, pairing the $RR$ sector of $\mathcal{S}$ with the $RL$ sector of $\mathcal{R}$.

\subsubsection{Boundary terms}

After the interior cancellations \eqref{eq:cancel_010}--\eqref{eq:cancel_diag} remove the $(1,2)$, $(+,2)$, $(1,-)$ and $(1,1)$ pairs, four $(a,b)$ endpoint pairs are left with nonzero total coefficient. These are the boundary terms where $\SfI$ reached the {\em end of the fermion line}
\begin{equation}
  \sum_X\mathcal{M}^P_X\big|_{\text{boundary}}
  = - (2,2)_{\cal L} +(+,1)_{\cal R}+(2,-)_{\cal L}-(+,-)_{\cal B}.
  \label{eq:surviving}
\end{equation}
The single surviving $(2,2)$ contribution is the residue from $\mathcal{L}_{RL}$ after the two $\mathcal{S}$ amplitude $(2,2)$ terms ($\mathcal{S}_{LL}$ and $\mathcal{S}_{RL}$, the loop at $\epsilon_2$ and the trapped pinch) and the $(1,1)$ terms ($\mathcal{S}_{RR}$ and $\mathcal{R}_{RL}$) have all canceled within their pairs.

In every surviving boundary term except $(+,-)$, the intermediate-line propagator $\Sf(q)$ is {\em unshifted}, the four nonzero pairs being $(2,2)$ ($q$-independent tree), $(+,1)$ and $(2,-)$ (single shifts on $p$ and on $p'$) and $(+,-)$ (doubly-shifted tree). This confirms that the interior cancellations have removed all gauge-dependent contributions to $\hat\Sigma(q)$; the boundary terms affect only the external legs. The pairs $(+,1)$, $(2,-)$ and $(+,-)$ each carry a collapse point on a fermion endpoint, so if the external fermions are placed on-shell they vanish identically through
\begin{equation} \label{Diraceq}
\SfI(p)\,u(p)=0 \qquad \bar u(p')\,\SfI(p')=0.
\end{equation}
We retain them here because the present Compton calculation is kept fully off-shell; they are disposed of in Section~\ref{sec:general_n}.

In the chain-walk picture, the four survivors are the contributions left at the ends of the chain, namely the $q$-independent tree $(2,2)$, the single-boundary shifts $(+,1)$ and $(2,-)$ and the doubly-shifted tree $(+,-)$ from the maximal topology.

The mechanism by which each of the four survivors is removed (the on-shell subtraction of the renormalized mass for the $q$-independent $(2,2)$ residue, external self-energies on the legs for the single-boundary shifts and the crossed ($u$-channel) amplitude for the doubly-shifted tree) is structurally identical to the general-$n$ boundary cancellation developed in Sec.~\ref{sec:general_n}, so we defer the explicit analysis to that broader setting (Sec.~\ref{sec:chain}). Removal of the surviving $q$-independent $(2,2)$ residue is then carried out in Sec.~\ref{sec:renorm_mass}.


\section{Amplitude with More External Photons}
\label{sec:general_n}

The calculation of the previous section demonstrated the cancellation mechanism for two external photon vertices. The extension to arbitrary $n$ rests on a single statement in the endpoint notation, namely that at every {\em internal} endpoint pair $(a,b)$ the four contributing diagrams sum to zero.

On a single open fermion line\footnote{A pinch photon connects two points on fermion lines. Insertions on two different open lines factorize into two single-line pinches, and an insertion on a closed fermion loop fires the WTI inside the cyclic trace, leaving no boundary term, so the single open fermion line is the universal case.} with $n+1$ external photon vertices, the number $k$ of photons between the two pinch-photon endpoints fixes the topology, with $k=0$ the self-energy, $k=1$ the triangle (vertex correction), $k=2$ the box, $k=3$ the pentagon and so on, up to $k=n+1$, the maximal box with both endpoints on the external legs.

For $n>1$, topologies with $k\geq 3$ external photons between the pinch-photon endpoints arise (pentagons, hexagons, etc.), and after the WTI fires at both endpoints each topology produces shifted-boundary terms that do not cancel within that single topology. They cancel {\em across adjacent topologies}, a single-boundary-shift term of topology $k$ sharing its Dirac structure with the neighbors $k{\pm}1$. We call this intuitive picture {\em $k$-telescoping}; the {\em $(a,b)$ sum rule} derived below makes it precise, organizing the cancellation by the post-pinch endpoint pair, and is the principle that closes the general-$n$ case. The $k$-telescoping is invisible in the Compton case ($n=1$), where the small number of topologies makes the inter-topology cancellations appear as isolated pairwise matchings.

\subsection{Setup}
\label{sec:four_term}

We keep the conventions of Section~\ref{sec:segments}, the $n+2$ segments indexed $s = 0, 1, \ldots, n+1$ with segment $s$ between $\gamma_s$ and $\gamma_{s+1}$ (and $\gamma_0 \equiv +$, $\gamma_{n+2} \equiv -$), the square bracket $[a,b]$ for a pre-pinch placement and the parentheses $(a,b)$ for a post-pinch endpoint pair. The placement amplitude~\eqref{eq:placement_amp}, its placement-to-endpoint map~\eqref{eq:placement_to_ab} and the post-pinch amplitude~\eqref{eq:ab_amp} set up there organize the cancellation that follows.

The WTI fires at each insertion, the L/R choice collapses each endpoint to one of its segment's bounding vertices, and the following four sectors produce diagrams with post-pinch endpoint pair
\begin{equation}
  [s_1, s_2] \to 
  \begin{cases}
   + (s_1+1,  s_2+1) & (LL) \\
   - (s_1,  s_2+1) & (LR )\\
   - (s_1+1,  s_2) & (RL )\\
   + (s_1,  s_2) & (RR)
  \end{cases}
  \label{eq:placement_to_ab}
\end{equation}
carrying the universal sign pattern $(+,-,-,+)$. 

\subsection{Cancellation}
\label{sec:dirac_identity}

Inverting~\eqref{eq:placement_to_ab}, a fixed endpoint pair $(a, b)$ with $a < b$ receives contributions from {\em four} placements,
\begin{equation}
\begin{array}{ll}
  \text{LL of placement } [a-1,  b-1] & \text{sign } +1 \\
  \text{LR of placement } [a,  b-1] & \text{sign } -1  \\
  \text{RL of placement } [a-1,  b] & \text{sign } -1  \\
  \text{RR of placement } [a,  b] & \text{sign } +1 
\end{array}
\label{eq:four_sources}
\end{equation}
all sharing the same integrand $(a,b)$ of~\eqref{eq:ab_amp} (propagators between $a$ and $b$ shifted, all others unshifted, with the common loop factor $1/\ell^4$ and the same Dirac structure). When all four placements lie in the allowed range $0 \le s_1 \le s_2 \le n+1$, they contribute the same amplitude $\mathcal{M}^P_{(a,b)}$ with signs $(+,-,-,+)$ we have seen in Section~\ref{sec:segments},
\begin{equation}
  +(a,b)_{LL} - (a,b)_{LR} - (a,b)_{RL} + (a,b)_{RR} = 0,
  \label{eq:ab_sum_rule}
\end{equation}
and the $(a,b)$ contribution cancels exactly. This is the {\em interior $(a,b)$ sum rule}, which applies whenever both $a, b \in \{1, \ldots, n+1\}$ are external photon vertices with $a < b$ and absorbs the entire ``$k$-telescoping'' picture into a single algebraic statement.

\begin{figure}[ht]
\centering
\begin{tikzpicture}[
  fermion/.style={thick, postaction={decorate,
    decoration={markings, mark=at position 0.50 with {\arrow{>}}}}},
  photon/.style={decorate, decoration={snake, amplitude=0.8pt,
    segment length=3pt}},
  pinchedphoton/.style={densely dotted, thick},
  vertex/.style={fill, circle, inner sep=0.9pt},
  scale=0.95, every node/.style={font=\scriptsize}
]
\def\drawbase{
  \draw[fermion] (6.2,0) node[right]{$p$} -- (0,0) node[left]{$p'$};
  \foreach \x in {0.7,1.4,2.1,2.8,3.5,4.2,4.9,5.6} \node[vertex] at (\x,0){};
  \foreach \x in {0.7,1.4,2.1,2.8,3.5,4.2,4.9,5.6} \draw[photon] (\x,0)--(\x,-0.55);
}
\begin{scope}[shift={(3.8,4.0)}]
  \node at (3.1,2.0) {placement $[2,6]$ (pre-pinch)};
  \drawbase
  \draw[photon] (1.75,0) .. controls (2.05,1.45) and (4.25,1.45) .. (4.55,0);
  \node at (3.15,1.4) {$\ell$};
  \node[vertex] at (1.75,0){};
    \node[vertex] at (4.55,0){};
\end{scope}
\begin{scope}[shift={(0,0)}]
  \node at (3.1,2.0) {$(a,b)_{LL}=(3,7)$};
  \drawbase
  \draw[pinchedphoton] (1.4,0) .. controls (1.6,1.35) and (4.0,1.35) .. (4.2,0);
  \node at (2.8,1.4) {$\ell$};
\end{scope}
\begin{scope}[shift={(7.6,0)}]
  \node at (3.1,2.0) {$(a,b)_{LR}=(2,7)$};
  \drawbase
  \draw[pinchedphoton] (1.4,0) .. controls (1.6,1.5) and (4.7,1.5) .. (4.9,0);
  \node at (3.15,1.4) {$\ell$};
\end{scope}
\begin{scope}[shift={(0,-3.7)}]
  \node at (3.1,2.0) {$(a,b)_{RL}=(3,6)$};
  \drawbase
  \draw[pinchedphoton] (2.1,0) .. controls (2.3,1.25) and (4.0,1.25) .. (4.2,0);
  \node at (3.15,1.4) {$\ell$};
\end{scope}
\begin{scope}[shift={(7.6,-3.7)}]
  \node at (3.1,2.0) {$(a,b)_{RR}=(2,6)$};
  \drawbase
  \draw[pinchedphoton] (2.1,0) .. controls (2.3,1.35) and (4.7,1.35) .. (4.9,0);
  \node at (3.5,1.4) {$\ell$};
\end{scope}
\end{tikzpicture}
\caption{Top, the pre-pinch placement $[s_1,s_2]=[2,6]$ in an $n=7$ fermion line, with the virtual photon attaching at the midpoints of segments $2$ and $6$ (the $\gamma_2\gamma_3$ and $\gamma_6\gamma_7$ intervals), which are not vertices. Bottom, the four sectors $LL, LR, RL, RR$ of this placement, drawn in the after-pinching picture. The two pinch insertions on segments $2$ and $6$ each fire L or R via the WTI, sending the pinch-photon endpoints to one of $\{\gamma_2, \gamma_3\}$ on the right and one of $\{\gamma_6, \gamma_7\}$ on the left; the four combinations produce four different endpoint pairs $(a,b)$. With the universal sign pattern $(+,-,-,+)$, this single placement contributes once to each of the four $(a,b)$ pairs $(2,6), (2,7), (3,6), (3,7)$.}
\label{fig:fourterm}
\end{figure}
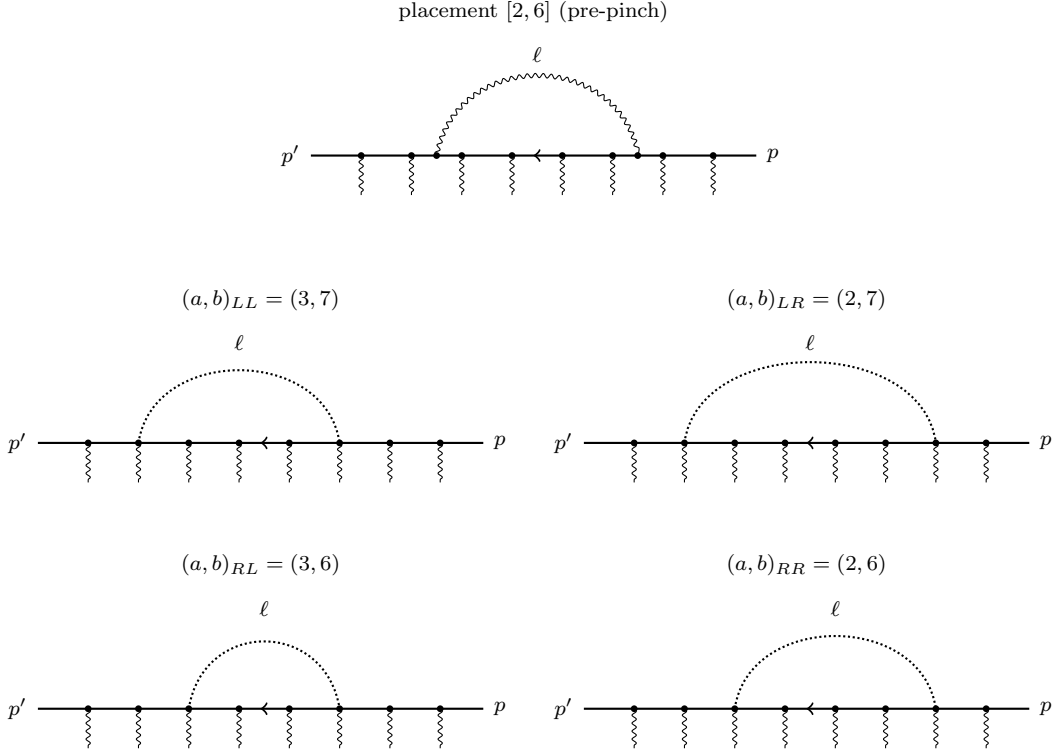

As an example, the four panels in Fig.~\ref{fig:fourterm} illustrate the four sectors at a single representative placement for $n=7$, showing how one placement $[s_1, s_2] = [2,6]$ distributes its four sectors to four different $(a,b)$ pairs $\{(2,6), (2,7), (3,6), (3,7)\}$. The $(a,b)$ sum rule \eqref{eq:ab_sum_rule} is the dual statement that at each fixed $(a,b)$ pair, the four placements that funnel into it via \eqref{eq:placement_to_ab} sum to zero.

\subsection{Surviving boundary terms}
\label{sec:chain}

The sum rule~\eqref{eq:ab_sum_rule} fails to cancel an $(a,b)$ contribution exactly when one or more of the four placements in~\eqref{eq:four_sources} falls outside the allowed range $0 \le s_1 \le s_2 \le n+1$. This happens precisely when at least one of $a, b$ is a fermion endpoint, or when $a = b$
\begin{equation}
\begin{array}{lll}
  (a, -),\ a \in \{1,\ldots,n+1\}: & \text{2 of 4 placements valid (LL, LR)}, & \text{net } 0,\\
  (+, b),\ b \in \{1,\ldots,n+1\}: & \text{2 of 4 placements valid (LR, RR)}, & \text{net } 0,\\
  (+, -): & \text{1 placement (LR at $[0,n+1]$)}, & \text{net } -1,\\
  (a, a),\ a \in \{1,\ldots,n+1\}: & \text{trapped (LL $[a{-}1, a{-}1]$, RL $[a{-}1, a]$, RR $[a,a]$)},
    & \text{net } -1
\end{array}
\label{eq:boundary_summary}
\end{equation}
where the trapped count for $(a,a)$ assumes the convention of Fig.~\ref{fig:pinched} placing the trapped-pinch residue at the $\gamma_a$ endpoint; the $(a,a)$ residues from same-segment placements combine to a single $-(a,a)$ contribution after the $\slashed\ell$ piece of the trapped pinch is dropped under symmetric integration. This surviving $-(a,a)$ is $q$-independent (a constant times the tree) and is disposed of not here but in the renormalized mass (Sec.~\ref{sec:renorm_mass}); the endpoint diagonals $(+,+) = [0,0]$ and $(-,-) = [n+1,n+1]$ are the leg analogs, likewise $q$-independent and absorbed by external-leg renormalization.

We now take the two outermost fermion legs, of momenta $p$ and $p'$, on-shell. The gauge-invariant self-energy $\hat\Sigma(q)$ depends only on the internal momentum $q$ (Section~\ref{sec:genSE}), so this restricts neither $\hat\Sigma$ nor the mass function, while through the Dirac equation (\ref{Diraceq}) it sends to zero every amplitude that carries a collapse point at a fermion endpoint. The boundary pairs $(+,b)$ and $(a,-)$, the doubly-shifted tree $(+,-)$ and the endpoint diagonals $(+,+)$ and $(-,-)$ then vanish identically, and the sole survivors of~\eqref{eq:boundary_summary} are the interior diagonals $(a,a)$. These are $q$-independent (each a constant times the tree Dirac structure, with no dependence on the line momentum), so although they survive the pinch cancellation they do not enter the renormalized mass function, since the on-shell subtraction of Section~\ref{sec:renorm_mass} removes any $q$-independent piece identically.

\paragraph{Off-shell boundary.} 
The interior/boundary classification follows directly from the segment picture. When a firing exhausts its segment, that is, when one of the two propagators in the difference is the endpoint propagator $\Sf(q_0)$ or $\Sf(q_n)$, the inverse propagator $\SfI$ finds no further propagator to cancel against, and a {\em boundary term} survives, in which the endpoint propagator is shifted, $\Sf(q_0)\to\Sf(q_0-\ell)$ or $\Sf(q_n)\to\Sf(q_n-\ell)$. When the firing does not reach the endpoint (both propagators in the difference are interior to the segment), the two terms cancel pairwise with corresponding terms from the other topologies. These are the {\em interior cancellations}. The decomposition is {\em universal}, applying identically regardless of how many photon vertices sit on the fermion line, and regardless of whether the external fermions are on-shell or off-shell. The only difference is that on-shell the boundary terms vanish automatically due to the Dirac equation (\ref{Diraceq}), whereas off-shell they must be canceled explicitly.

Keeping the outermost legs off-shell, as in the explicit Compton calculation of Section~\ref{sec:explicit}, leaves these endpoint amplitudes present, but they need not be removed by hand. Each carries a collapse point on a fermion endpoint and shifts only an external-leg propagator, so none of them touches the intermediate line $\Sf(q)$ or enters $\hat\Sigma$. An off-shell fermion leg is moreover never a true boundary, since it is itself an internal line of some larger diagram whose genuine external legs are on-shell; the WTI telescoping that generated the boundary term then continues across the apparent endpoint and terminates only at the on-shell asymptotic states, where $\SfI(p)\,u(p)=0$ and $\bar u(p')\,\SfI(p')=0$ remove it. Taking the two outermost legs on-shell is therefore without loss of generality rather than a physical restriction, and the explicit off-shell cancellation of the Compton case (the boundary pairs $(+,b)$ and $(a,-)$ against the external-leg self-energies, the doubly-shifted tree $(+,-)$ against the $u$-channel) is a concrete instance of it rather than an additional requirement. Either way $\hat\Sigma(q)$ is the same, since it depends only on~$q$.

\paragraph{Total cancellation for fixed $n$.} Aggregating across all placements $[s_1, s_2]$ with $0 \le s_1 \le s_2 \le n+1$, the four-source sum rule~\eqref{eq:ab_sum_rule} clears every interior $(a, b)$ pair, and the partial sum rule clears the boundary pairs $(+, b)$ and $(a, -)$, leaving only the diagonal residues and the doubly-shifted tree
\begin{equation}
  \sum_{0 \le s_1 \le s_2 \le n+1}
    \mathcal{M}^P_{[s_1, s_2]}
  \;=\;
  -\sum_{a=1}^{n+1} \mathcal{M}^P_{(a,a)}
   \,-\, \mathcal{M}^P_{(+,-)}.
  \label{eq:surviving_general_n}
\end{equation}
Under the on-shell choice above the doubly-shifted tree $\mathcal{M}^P_{(+,-)}$ has already vanished, and only the $q$-independent $\mathcal{M}^P_{(a,a)}$ remains, removed by the on-shell subtraction that defines the renormalized mass (Sec.~\ref{sec:renorm_mass}); off-shell, $\mathcal{M}^P_{(+,-)}$ is removed instead by the crossed ($u$-channel) amplitude. Combining,
\begin{equation}
  \sum_{[s_1, s_2]} \mathcal{M}^P_{[s_1, s_2]}
  \;+\; \mathcal{M}^P_u \;+\; \delta_\text{c.t.}
  \;=\; 0,
  \label{eq:total_cancellation}
\end{equation}
the sum of all one-loop $D^P$ contributions vanishes for every fixed $n$. This is the general-$n$ extension of the explicit Compton ($n=1$) cancellation~\eqref{eq:surviving}.

Specialized to $n=1$, Eq.~\eqref{eq:surviving_general_n} gives $-\mathcal{M}^P_{(1,1)} - \mathcal{M}^P_{(2,2)} - \mathcal{M}^P_{(+,-)}$. The bookkeeping of Sec.~\ref{sec:cancel} separates external SE on the legs from the four $s$-channel topologies, redistributing this content into the three interior sum rules at $(1, 2)$, $(+, 2)$, $(1, -)$ and the four surviving boundary terms $(2, 2), (+, 1), (2, -), (+, -)$ of Eq.~\eqref{eq:surviving}; the two bookkeepings are equivalent, and both reduce to~\eqref{eq:ab_sum_rule} at their respective endpoint pairs.


\section{The Generalized Self-Energy and Vertex}
\label{sec:genSE}

The cancellation analysis of Section~\ref{sec:cancel} sorted the $(1-\xi)$ contributions of the four pinched amplitudes into interior pieces (which cancel) and boundary survivors (which cancel against the external SE pinches and the $u$-channel). The same four-sector split feeds two gauge-invariant objects, with the propagator-like contributions packaging into a self-energy $\hat\Sigma$ on the intermediate fermion line and the complementary vertex-like contributions into a modified vertex $\hat\Gamma$. We construct both here.

\subsection{The generalized self-energy}

The propagator-like piece $\Delta\Sigma^{(X)}$ of each pinched amplitude $X \in \{\mathcal{R},\mathcal{L},\mathcal{B}\}$ is the part whose Dirac structure differs from the tree amplitude only through $\Sf(q)\to\Sf(q-\ell)$, with the external-leg propagators $\Sf(p)$ and $\Sf(p')$ untouched. Reading off the amplitudes of Section~\ref{sec:amplitudes}, these are the sector sums
\begin{equation}
  \Delta\Sigma^{(\mathcal{R})} = \mathcal{R}_{LL}+\mathcal{R}_{RL},
  \quad
  \Delta\Sigma^{(\mathcal{L})} = \mathcal{L}_{RL}+\mathcal{L}_{RR},
  \quad
  \Delta\Sigma^{(\mathcal{B})} = \mathcal{B}_{RL},
\end{equation}
each evaluating to $\Sf(p')\,\slashed\epsilon_2^*\,[\Sf(q-\ell)-\Sf(q)]\,\slashed\epsilon_1\,\Sf(p)$, with $\Sf(q-\ell)$ alone replacing the difference for $\mathcal{B}$ and the external legs unshifted. The reference half $-\Sf(q)$ of each difference has a $q$-independent loop integral and is the same residue that Section~\ref{sec:cancel} lists as the surviving diagonal $(2,2)$ term, so whether it is grouped into $\hat\Sigma$ here or left as a boundary survivor there makes no difference, since the on-shell subtraction of Section~\ref{sec:renorm_mass} removes it either way. The $\mathcal{S}$ amplitude is already the self-energy topology and contributes through the conventional $\Sigma_\xi(q)$.

The {\em generalized gauge-invariant self-energy} absorbs them all
\begin{equation}
  \hat\Sigma(q) = \Sigma_\xi(q)
  + \Delta\Sigma^{(\mathcal{R})}(q;p)
  + \Delta\Sigma^{(\mathcal{L})}(q;p')
  + \Delta\Sigma^{(\mathcal{B})}(q;p,p').
  \label{eq:hatSigma}
\end{equation}
The $(1-\xi)$ piece $\Sigma_P(q)$ of $\Sigma_\xi$, eq.~\eqref{SigmaP}, cancels against the sum of $\Delta\Sigma$'s by the interior cancellation~\eqref{eq:cancel_010}, leaving
\begin{equation}
  \hat\Sigma(q) = \Sigma_\xi(q)\big|_{\xi=1}.
  \label{eq:hatSigma_Feynman}
\end{equation}
Although each $\Delta\Sigma^{(X)}$ was extracted from a diagram whose integrand involves the external momenta, the total $\hat\Sigma(q)$ depends only on~$q$, and $\partial\hat\Sigma/\partial\xi = 0$ is immediate. Process independence is manifest, since Compton scattering and pair annihilation $e^+e^-\to\gamma\gamma$, related by crossing $p'\to -p'$, yield the same $\hat\Sigma(q)$. The derivation never uses the on-shell condition or transversality of the external photons, so the construction applies equally for off-shell external photons (form factors, higher-loop embeddings, processes where the photon connects to another fermion line).

The same four local topologies build $\hat\Sigma$ on any internal segment for arbitrary $n$. The self-energy on segment $s$ is the placement $[s,s]$, with both pinch insertions on that segment, and its gauge-parameter-dependent part $\Sigma_P$ is the propagator-like piece that must cancel, the shift of the segment-$s$ propagator carried by the endpoint pair $(s,s+1)$. Under the sum rule~\eqref{eq:ab_sum_rule} this pair has four sources, the self-energy $[s,s]$ together with the two vertices $[s,s+1]$ and $[s-1,s]$ and the box $[s-1,s+1]$, the direct generalization of the Compton set $\mathcal{S},\mathcal{R},\mathcal{L},\mathcal{B}$; their gauge-dependent $(s,s+1)$ contributions cancel, leaving the gauge-invariant $\hat\Sigma=\Sigma_\xi(q)\big|_{\xi=1}$. A pentagon or higher topology has no single-segment sector, so it never reaches $(s,s+1)$, and the box is the largest topology that enters $\hat\Sigma$ at any $n$.

Clearing $(s,s+1)$ is necessary but is not the whole cancellation. The same four placements do not fire into $(s,s+1)$ alone; each also deposits gauge-parameter-dependent boundary terms on the neighboring endpoint pairs, and removing those recruits the next topologies outward, so the cancellation telescopes segment by segment. The full-amplitude gauge cancellation therefore requires arbitrarily high topologies (Section~\ref{sec:general_n}), and every diagram is needed for it, even though only those through the box ever reach $\hat\Sigma$ itself.

The pair $(s,s+1)$ places one endpoint at each of the two vertices that bound segment $s$. The self-energy on a segment is therefore assembled from the firing at the vertex on each side, half from the left bounding vertex and half from the right, since each interior vertex is shared by the two segments it separates and splits its firing evenly between them. This is the dual of the shared-vertex reading of the modified vertex, where a vertex sitting between two segments draws half its longitudinal content from the self-energy on each side.

This is a known result of the pinch technique~\cite{Cornwall:Papavassiliou:1989,Papavassiliou:1990,Denner:Dittmaier:Weiglein:1994}. The present derivation establishes it more generally in two respects: it follows from the segment-local WTI alone, without ever invoking the Dirac equation or on-shell external states, and it holds on every internal segment of a fermion line with an arbitrary number of external vertices, the same four local topologies building $\hat\Sigma$ for any $n$. Uniqueness of the propagator-like classification is established in the uniqueness analysis below.

\subsection{The gauge-parameter invariant vertex $\hat\Gamma$}

The same decomposition that defines $\hat\Sigma$ simultaneously defines the gauge-parameter invariant vertex. In each topology, the pinched amplitude $X$ splits into propagator-like terms that are absorbed into $\hat\Sigma$ and the complementary vertex-like terms absorbed into $\hat\Gamma$.

\subsubsection*{Right vertex}

Applying the fundamental identity \eqref{eq:fund_intro} at both pinch insertions of $\Gamma_P^\nu$ in \eqref{eq:MR}, with $D^P_{\rho\sigma}(\ell) = (1-\xi)i\ell^\rho\ell^\sigma/\ell^4$, the $\ell^\rho,\ell^\sigma$ contract with the adjacent $\gamma^\rho, \gamma^\sigma$ to produce a $\slashed\ell$ at each loop endpoint, and each $\slashed\ell$ sits between two propagators ready for the identity to fire. The result is the manifestly factorized form
\begin{equation}
  \Sf(q)\,\Gamma_P^\nu(q,p)\, \epsilon_{1,\nu}\,\Sf(p)
  = -i(1-\xi)g^2 C_2 \dL \frac{1}{\ell^4}\,
     \bigl[\Sf(q)-\Sf(q-\ell)\bigr]\slashed\epsilon_1
     \bigl[\Sf(p-\ell)-\Sf(p)\bigr],
  \label{eq:GammaR}
\end{equation}
in which each bracket factor is the propagator difference produced by one of the two pinches firing. Expanding the product yields the four sectors $\mathcal{R}^{IJ}$ with the universal $(+,-,-,+)$ sign pattern. Two of them with $\Sf(p)$ are the propagator-like pair $\Delta\Sigma^{(\mathcal{R})}$ already absorbed into $\hat\Sigma$; the two terms with $\Sf(p-\ell)$ surviving carry the shifted $p$-leg propagator, the hallmark of a vertex correction at the $\epsilon_1$-vertex
\begin{equation}
  \label{eq:DeltaGammaR}
  \Delta\Gamma^{(\mathcal{R})}
  = \mathcal{R}_{LR}+\mathcal{R}_{RR} 
  =   -i(1-\xi)\, e^2 g^2 C_2 \dL \frac{1}{\ell^4}\,
    \Sf(p')\slashed\epsilon_2^*
    \bigl[\Sf(q)-\Sf(q-\ell)\bigr]\slashed\epsilon_1\,\Sf(p-\ell),
\end{equation}
the integrand differing from the tree amplitude only through $\Sf(q)\to[\Sf(q)-\Sf(q-\ell)]$ at the intermediate line and $\Sf(p)\to\Sf(p-\ell)$ at the $p$-leg.

\subsubsection*{Left vertex}

Analogously, applying \eqref{eq:fund_intro} at both pinch insertions of $\bar\Gamma_P^\mu$ in \eqref{eq:ML} gives
\begin{equation}
  \Sf(p')\,\bar\Gamma_P^\mu(p',q)\, \epsilon_{2,\mu}^*\,\Sf(q)
  = -i(1-\xi)g^2 C_2 \dL \frac{1}{\ell^4}\,
     \bigl[\Sf(p'-\ell)-\Sf(p')\bigr]\slashed\epsilon_2^*
     \bigl[\Sf(q)-\Sf(q-\ell)\bigr].
  \label{eq:GammaL}
\end{equation}
Expanding the left bracket separates the $\mathcal{L}^{IJ}$ sectors into the propagator-like pair $\Delta\Sigma^{(\mathcal{L})}$ (absorbed into $\hat\Sigma$) and the vertex-like pair. The latter carries the shifted $p'$-leg, the hallmark of a vertex correction at the $\epsilon_2^*$-vertex
\begin{equation}    \label{eq:DeltaGammaL}
  \Delta\Gamma^{(\mathcal{L})} =\mathcal{L}_{LL}+\mathcal{L}_{LR}
   =  -i(1-\xi)\, e^2 g^2 C_2 \dL \frac{1}{\ell^4}\,
    \Sf(p'-\ell)\slashed\epsilon_2^*
    \bigl[\Sf(q)-\Sf(q-\ell)\bigr]\slashed\epsilon_1\,\Sf(p),
\end{equation}
the integrand differing from the tree amplitude through $\Sf(p')\to\Sf(p'-\ell)$ at the $p'$-leg and $\Sf(q)\to[\Sf(q)-\Sf(q-\ell)]$ at the intermediate line.

\subsubsection*{Box}

From $\mathcal{B}$, after the propagator-like $\mathcal{B}_{RL}=\Delta\Sigma^{(\mathcal{B})}$ is absorbed into $\hat\Sigma$, we are left with $\mathcal{B}_{LL}$, $\mathcal{B}_{LR}$ and $\mathcal{B}_{RR}$. All three carry at least one shifted external-leg propagator, with the intermediate propagator always shifted. These define the modified box contribution $\Delta\mathcal{B}$.

\subsubsection*{Structure of the decomposition}

Note that the intermediate-line propagator difference in \eqref{eq:DeltaGammaR} and \eqref{eq:DeltaGammaL} is $[\Sf(q)-\Sf(q-\ell)]$, the {\em opposite sign} of the difference $[\Sf(q-\ell)-\Sf(q)]$ that entered $\hat\Sigma$. This is not accidental; the WTI at the vertex adjacent to the intermediate line produces $\Sf(q-\ell)-\Sf(q)$, and the two signs partition into $\hat\Sigma$ and $\hat\Gamma$ with a relative minus sign that ensures the amplitude $X$ is reconstructed exactly
\begin{equation}
  X
  = \Delta\Sigma^{(X)}
  + \Delta\Gamma^{(X)}
  \qquad
  \text{for each }X\in\{\mathcal{R},\mathcal{L},
  \mathcal{B}\}.
\end{equation}
The modified vertex and self-energy together account for all pinch contributions; nothing is lost or double-counted.

Since $\Delta\Sigma^{(X)}$ is proportional to $(1-\xi)$, so is $\Delta\Gamma^{(X)}$, since both originate from $D^P_{\mu\nu}$. The modified vertex is therefore
\begin{equation}
  \hat\Gamma^\mu
  = \Gamma_{\xi}^\mu
  + \Delta\Gamma^{(\text{pinch}),\mu},
\end{equation}
and its gauge invariance follows from the same mechanism as for $\hat\Sigma$, with the pinch contributions cancelling the $\xi$-dependent part of the conventional vertex to yield $\hat\Gamma^\mu = \Gamma_{\xi}^\mu\big|_{\xi=1}$.

\subsubsection*{The transverse vertex}

It is the full vertex, not only its longitudinal projection, that is made gauge invariant. Contracting the gauge-dependent part of the conventional vertex at the $\epsilon_1$ vertex with the photon momentum $q-p$ returns, by the Ward--Takahashi identity~\eqref{eq:exact_WTI},
\begin{equation}
  (q-p)_\nu\,\Gamma^\nu_\xi(q,p)\big|_{(1-\xi)\text{ part}}
  = -\bigl[\Sigma_P(q)-\Sigma_P(p)\bigr],
  \label{eq:vertex_long}
\end{equation}
a relation that fixes only the longitudinal projection. The transverse vertex obeys $(q-p)_\nu\Gamma^\nu_T=0$ identically, so the identity is silent on it. Read in isolation this suggests the transverse vertex is an independent object the construction never reaches. It is not.

The gauge-dependent content of the vertex is not a longitudinal object. It is generated by the same $D^P$ photon and the same identity~\eqref{eq:fund_intro} that build $\Sigma_P(q)=(1-\xi)(\slashed q-m)\Omega(q)$, and it inherits their full Dirac and Lorentz structure: the factorized forms~\eqref{eq:DeltaGammaR} and~\eqref{eq:DeltaGammaL}, together with the boundary terms proportional to $(\slashed q-m)$ and $(\slashed p-m)$ at the two endpoints, all carry a nonzero transverse projection. The contraction in~\eqref{eq:vertex_long} sees only the longitudinal combination, the difference $\Sigma_P(q)-\Sigma_P(p)$; the value of $\Sigma_P$ at each endpoint, which feeds the transverse part as well, it cannot see.

The pinch contributions cancel this content in full. The cancellation acts on the whole tensor, not on the longitudinal difference alone, so it removes the transverse gauge dependence together with the longitudinal; the surviving boundary terms are taken up by the external-leg self-energies and the $u$-channel, and $\hat\Gamma^\mu=\Gamma^\mu_\xi|_{\xi=1}$ holds as a full tensor identity rather than in its longitudinal projection alone. The modification is genuinely off shell: on shell the boundary terms vanish, $\Sigma_P$ drops out and the transverse vertex is already gauge invariant, which is why the question does not arise in the on-shell pinch technique. This off-shell robustness has the same origin as for $\hat\Sigma$: the gauge-dependent pieces cancel through the four-sector sum rule~\eqref{eq:ab_sum_rule} with its universal $(+,-,-,+)$ signs, all four sectors kept distinct because none collapses under a Dirac equation. Gauge invariance of $\hat\Gamma^\mu_T$ is the removal of its gauge dependence, not a determination of its value, which the $\xi=1$ calculation supplies; the mass function $m(q)$ depends only on $\hat\Sigma$ and is unaffected either way.

\subsection{Uniqueness of the decomposition}
\label{sec:uniqueness}

We now clarify the distinct roles of segment locality (Section~\ref{sec:segments}), the structural classification and gauge independence. Segment locality (each WTI firing telescopes entirely within its own segment, with the segment boundaries inert) ensures that the two pinch endpoints act independently. The combined output of the two firings gives the four terms of the amplitude.

\textit{The structural criterion} classifies these four terms, a term being propagator-like if and only if the external-leg propagators $\Sf(p)$, $\Sf(p')$ are unshifted, with only $\Sf(q)\to\Sf(q-\ell)$ on the intermediate line. This criterion is unambiguous (there is no freedom in which propagators carry a shifted momentum) and it assigns each of the four terms to either $\Delta\Sigma^{(X)}$ or $\Delta\Gamma^{(X)}$ without residual choice.

\textit{The nontrivial content} is that this criterion perfectly separates the $(1-\xi)$ contributions, since all and only the $(1-\xi)$ pieces that modify the intermediate-line propagator satisfy the unshifted-external-legs criterion. This alignment is not guaranteed \textit{a priori}; it is a consequence of the specific way the WTI distributes $(1-\xi)$ contributions across the four topologies. Were there a $(1-\xi)$ term that shifted an external leg while leaving the intermediate line alone, it would be classified as vertex-like despite being gauge-dependent, and the clean separation would fail. It does not fail, because the same WTI that generates the gauge dependence is the identity the pinch procedure uses to redistribute it.

\subsubsection*{Uniqueness from gauge independence}

In principle one could absorb only a fraction $\alpha$ of the structurally identified terms into the self-energy and leave the remaining $(1-\alpha)$ in the vertex, defining a one-parameter family $\hat\Sigma_\alpha = \Sigma_\xi|_{\xi=1} + (1-\alpha)\Sigma_P$. For every $\alpha$ the total amplitude is unchanged, the split being only a relabeling of propagator-like terms between the self-energy and the vertex. But for $\alpha\neq 1$ the residual $(1-\alpha)(1-\xi)$ piece makes $\hat\Sigma_\alpha$ depend on the starting gauge, contaminating the wave-function renormalization with a $\xi$-dependent factor $Z_\alpha\neq 1$. The requirement that $\hat\Sigma$ be intrinsic to the theory, the same answer regardless of which $R_\xi$ gauge one computes in, selects $\alpha=1$ uniquely. At $\alpha=1$ the entire $\xi$-dependent part is projected out, the dressed propagator takes the canonical form $i/(\slashed q - m(q))$ with unit residue on $\slashed q$, and the mass function is read off directly. An independent confirmation is the PT--BFM equivalence~\cite{Denner:Dittmaier:Weiglein:1994,Hashimoto:Kodaira:Yasui:Sasaki:1994,Papavassiliou:1995}, in which the background field method in Feynman gauge defines gauge-invariant Green's functions by an entirely different route and the result agrees. The $\alpha$ freedom is a genuinely off-shell layer of uniqueness with no counterpart in the standard PT~\cite{Cornwall:Papavassiliou:1989,Papavassiliou:1990,Binosi:Papavassiliou:2004}, where the Dirac equation on the external spinors eliminates boundary terms before the question of partial pinching can arise.

A distinct one-parameter freedom is the choice of reference gauge. Pinching the longitudinal content relative to an arbitrary $\xi_0$, rather than relative to Feynman gauge, removes the part proportional to $(\xi_0-\xi)$ and returns $\Sigma_{\xi_0}$, which is independent of the computational gauge $\xi$ but still carries the reference value through $(1-\xi_0)(\slashed q-m)\Omega(q)$. This is not the same as the partial-pinch family $\hat\Sigma_\alpha$ above: every member $\Sigma_{\xi_0}$ is independent of $\xi$, whereas $\hat\Sigma_\alpha$ is $\xi$-dependent for every $\alpha\neq1$; the two families coincide only at $\alpha=\xi_0=1$. Every member agrees on shell, since the members differ only by a term proportional to $(\slashed q-m)$, which is why the ambiguity does not arise in the on-shell pinch technique. Off shell the difference survives, after on-shell renormalization as $(1-\xi_0)(\slashed q-m)[\Omega(q)-\Omega(m)]$, a kinetic term that the WTI reshuffles into the corresponding vertex, so each $m_{\xi_0}(q)$ paired with its vertex gives the same amplitude and has the correct pole $m_{\xi_0}(m)=m$. Locality therefore selects no preferred $\xi_0$, and the family is physical, each $m_{\xi_0}$ being the mass function of the fermion in the corresponding dressing~\cite{Choi:2026yxc,Lavelle:McMullan:1997}. We adopt $\xi_0=1$ throughout as the canonical representative, the transverse fixed point with no longitudinal part left to remove.

\section{The Gauge-Invariant Off-Shell Mass}
\label{sec:mass}

We now use the gauge-invariant self-energy $\hat\Sigma(q)$, together with the invariant vertices $\hat \Gamma(k',k)$, to define the off-shell mass function and renormalize it on shell.

\subsection{The renormalized mass function and the dressed propagator}
\label{sec:renorm_mass}

The gauge-invariant self-energy $\hat\Sigma(q)$ defines a gauge-invariant, Dirac-valued mass function
\begin{equation}
  m_B + \hat\Sigma(q),
  \label{eq:Mdef}
\end{equation}
whose pole fixes the physical mass through the relation $[\slashed{q}-m_B-\hat\Sigma(q)]_{\slashed{q}=m}=0$. The corresponding propagator does not, however, have unit residue at the pole. On-shell renormalization makes the pole mass explicit and simultaneously disposes of the surviving $q$-independent $(a,a)$ residues of Section~\ref{sec:cancel}.

Define the renormalized gauge-invariant self-energy by the on-shell subtraction
\begin{equation}
  \hat\Sigma^{\text{ren}}(q)
  = \hat\Sigma(q) - \hat\Sigma(m) - (\slashed{q}-m) \hat\Sigma'(m),
  \label{eq:hatSigmaR}
\end{equation}
where $m$ is now the physical (pole) mass and $\hat\Sigma'(q)\equiv d\hat\Sigma/d\slashed{q}$. The {\em renormalized mass function} is defined by absorbing the entire renormalized self-energy into the mass
\begin{equation}
  m(q) = m + \hat\Sigma^{\text{ren}}(q).
  \label{eq:MRdef}
\end{equation}
The renormalized dressed propagator then takes the form
\begin{equation}
  \hat\Sf^{\text{ren}}(q)
  = \frac{i}{\slashed{q}-m(q)},
  \label{eq:dressedR}
\end{equation}
where the entire renormalized self-energy has been absorbed into the Dirac-valued mass function $m(q)$ of \eqref{eq:MRdef}, so that the kinetic operator $\slashed{q}$ is displayed with unit coefficient and no separate momentum-dependent wave-function factor multiplies the propagator.

Conventionally, the last term of the self-energy in (\ref{eq:hatSigmaR}) is a field-strength renormalization carried by a separate momentum-dependent factor. Requiring the dressed propagator to keep the canonical form \eqref{eq:dressedR}, with unit coefficient on $\slashed q$ and unit residue at the pole, leaves that part no separate factor to reside in, so it is carried by $m(q)$. The division of $\hat\Sigma^{\text{ren}}$ between mass and field strength is a convention, fixed here by canonical normalization, which assigns the whole of it to the mass. Because $\hat\Sigma^{\text{ren}}$ is gauge-invariant, the resulting $m(q)$ is unambiguously gauge-invariant, unlike the conventional self-energy in which the mass and field-strength pieces are separately gauge-dependent and the assignment is ambiguous.

The two on-shell conditions enforce
\begin{enumerate}
\item $\hat\Sigma^{\text{ren}}(m)=0$, which gives
        \begin{equation}
          m(q)\big|_{\slashed{q}=m} = m.
        \end{equation}
The mass function at the pole equals the physical mass by construction.
\item ${\hat\Sigma^{\text{ren}\prime}}(m)=0$, which gives ${m}'(m) = 0$. Near the pole, $\slashed{q}-m(q) \approx (\slashed{q}-m)(1-{m}'(m)) = \slashed{q}-m$, so the propagator has residue~$i$, the standard normalization, without any momentum-dependent wave-function factor.
\end{enumerate}

Since $\hat\Sigma(q) = \Sigma_\xi(q)|_{\xi=1}$, as shown in (\ref{eq:hatSigma_Feynman}), the subtraction constants $\hat\Sigma(m)$ and $\hat\Sigma'(m)$ are manifestly gauge-invariant; they are the conventional self-energy and its derivative evaluated at a specific momentum \emph{and} at $\xi=1$, with no residual $\xi$-dependence. The renormalized self-energy $\hat\Sigma^{\text{ren}}(q)$ is therefore gauge-invariant at every $q$, and so is the mass function
\begin{equation}
  \frac{\partial m}{\partial\xi} = 0
  \qquad\text{for all }q.
\end{equation}

We now prove the claim deferred from Section~\ref{sec:cancel}, that the surviving $q$-independent $(a,a)$ residues are removed by the on-shell subtraction. Define the on-shell counterterm
\begin{equation}
  \Sigma_{\text{ct}}(q)
  = -\Sigma(m) - (\slashed{q}-m) \Sigma'(m),
  \label{eq:ct}
\end{equation}
where $\Sigma'(q)\equiv d\Sigma/d\slashed{q}$, evaluated at $\slashed{q}=m$. The gauge-dependent part $\Sigma_P(q)=(1-\xi)(\slashed{q}-m)\Omega(q)$ contributes to the counterterm as follows

\begin{enumerate}
\item {\em Mass counterterm.} $\Sigma_P(m) = 0$, because the overall factor $(\slashed{q}-m)$ vanishes on-shell.

\item {\em Wave-function counterterm.} Differentiating $\Sigma_P(q) = (1-\xi)(\slashed{q}-m)\Omega(q)$ with respect to $\slashed{q}$ and evaluating at $\slashed{q}=m$, the term from differentiating the overall $(\slashed{q}-m)$ gives $\Omega(m)$, while the term from differentiating $\Omega$ is killed by $(\slashed{q}-m)|_m = 0$. Therefore $\Sigma_P'(m) = (1-\xi)\Omega(m)$.
\end{enumerate}

The renormalized gauge-dependent self-energy is thus
\begin{equation}
  \Sigma^{P,\text{ren}}(q)
  = (1-\xi)\,(\slashed{q}-m) 
    \bigl[\Omega(q)-\Omega(m)\bigr].
  \label{eq:SigmaRP}
\end{equation}
The $q^2$-independent part of $\Omega$ is precisely the piece that, in the amplitude, produces the $(a,a)$ residue of Section~\ref{sec:cancel} (both WTIs fire completely, all $\ell$-dependence drops out of the Dirac structure and the integrand reduces to the tree amplitude over $\ell^4$). Being $q^2$-independent, it cancels identically in the difference $\Omega(q)-\Omega(m)$; we never need to assign a value to the scaleless loop integral itself.

At the level of the amplitude, the counterterm insertion on the $p$-leg generates a gauge-dependent, tree-proportional contribution $-(1-\xi)\Omega(m)\times\mathcal{M}_0$ (the tree Compton amplitude), and similarly on the $p'$-leg. These counterterm contributions absorb exactly the $(a,a)$ residue, which is thus removed by the subtraction rather than by any scaleless condition. In the Feynman-gauge formulation, this is automatic, the $q$-independent gauge-dependent piece never appearing because $\hat\Sigma(q)$ is evaluated at $\xi=1$ from the start.

The renormalized mass function $m(q)$ is the natural object, in that it is gauge-invariant, process-independent, equals the physical mass at the pole ($m(m)=m$) and its construction does not rely on any regularization-specific identity. The propagator $\hat\Sf^{\text{ren}}(q)=i/(\slashed{q}-m(q))$ has unit residue and a strictly local kinetic term.

The structure of the on-shell subtraction \eqref{eq:hatSigmaR} also guarantees decoupling. The renormalized correction $\hat\Sigma^{\text{ren}}(q) = \hat\Sigma(q)-\hat\Sigma(m)-(\slashed{q}-m)\hat\Sigma'(m)$ vanishes at the pole by construction and, when a heavy particle of mass $M\gg m$ circulates in the loop, the contribution to $\hat\Sigma^{\text{ren}}(q)$ is suppressed as $(q^2-m^2)^2/M^2$ in accordance with the Appelquist--Carazzone theorem. This was shown explicitly in the scalar sector~\cite{Choi:2024cbs,Choi:2024hkd}; the same mechanism operates here, with the additional gauge-invariance enforced by the pinch technique ensuring that the decoupling is an unambiguous, gauge-independent statement.

\subsection{Infrared structure and the scalar mass function}
\label{sec:IR}

The on-shell subtraction \eqref{eq:hatSigmaR} inherits the infrared sensitivity of every on-shell wave-function renormalization in QED. With a massless photon the derivative $\hat\Sigma'(m)$ is infrared divergent; regulated with a small photon mass $\lambda$, it diverges logarithmically, $\hat\Sigma'(m) \propto \ln\lambda^2$. The subtraction constant $\hat\Sigma(m)$, and with it the pole mass, is infrared finite, so the divergence enters $m(q)$ only through the term linear in $(\slashed q - m)$, precisely the piece that the canonical normalization above assigns to the mass function rather than to a separate residue factor. At the pole $m(m)=m$ is untouched; away from it, $m(q)$ is defined at fixed infrared regulator. This is the standard infrared structure of the on-shell scheme. In any physical rate the same logarithm cancels against real soft-photon emission by the Bloch--Nordsieck mechanism and its Kinoshita--Lee--Nauenberg generalization~\cite{Bloch:Nordsieck:1937,Kinoshita:1962,Lee:1964is}, so predictions built from $m(q)$ are infrared safe even though the function itself, like the conventional on-shell $Z_2$, is not.

An infrared-finite companion follows from the same gauge-invariant self-energy. Decompose
\begin{equation}
  \hat\Sigma(q) = \hat A(q^2)\,\slashed q + \hat B(q^2),
  \label{eq:ABdecomp}
\end{equation}
each coefficient separately gauge-invariant because $\hat\Sigma$ is. The dressed propagator then reads
\begin{equation}
  \hat\Sf(q)
  = \frac{i}{\slashed q - m_B - \hat\Sigma(q)}
  = \frac{i\,\hat Z(q^2)}{\slashed q - \hat M(q^2)},
  \qquad
  \hat Z(q^2) = \frac{1}{1-\hat A(q^2)},
  \qquad
  \hat M(q^2) = \frac{m_B + \hat B(q^2)}{1-\hat A(q^2)},
  \label{eq:Mhat}
\end{equation}
which defines the scalar mass function $\hat M(q^2)$. Its ultraviolet divergence, being momentum-independent at one loop, is removed by the single mass renormalization that trades $m_B$ for the pole mass through $\hat M(m^2)=m$; no wave-function subtraction enters, because a multiplicative rescaling of the propagator shifts $1-\hat A$ and $m_B+\hat B$ by a common factor that cancels in the ratio. The combination is therefore multiplicatively renormalization-group invariant, infrared finite since no on-shell derivative is subtracted and gauge-invariant by the construction of Section~\ref{sec:genSE}. Gauge invariance is exactly the property that the conventional $M(q^2)=B/(1-A)$ of the Schwinger--Dyson and lattice literature~\cite{Roberts:Schmidt:2020,Zhang:2004}, computed in Landau gauge, lacks, so $\hat M(q^2)$ is the object directly comparable with those determinations.

The two mass functions describe the same propagator and agree at the pole. Exactly,
\begin{equation}
  \slashed q - m(q)
  = \bigl[1-\hat A(q^2)\bigr]\bigl[\slashed q - \hat M(q^2)\bigr]
  + (\slashed q - m)\,\hat\Sigma'(m),
  \label{eq:mM_relation}
\end{equation}
which follows from \eqref{eq:hatSigmaR}, \eqref{eq:MRdef} and \eqref{eq:Mhat} together with the pole condition $m = m_B + \hat\Sigma(m)$; both sides vanish at $\slashed q = m$ and their derivatives there equal unity, consistent with unit residue, and the momentum-independent ultraviolet constant of $\hat A$ cancels between the two terms on the right. The relation localizes the infrared divergence: it resides entirely in the last term, the residue subtraction, while $\hat M(q^2)$ is infrared finite. The Dirac-valued $m(q)$ realizes the canonical propagator form $i/(\slashed q - m(q))$ at the price of carrying the regulated residue logarithm; the scalar $\hat M(q^2)$ forgoes the canonical form, retaining the residue factor $\hat Z(q^2)$, and is free of infrared sensitivity. Either presentation carries the same gauge-invariant content, and the choice between them is one of normalization, not of physics.

\subsection{$m(q)$ as a conceptual extension of mass}

In the conventional formalism, the fermion mass is a single number, namely the pole of the propagator $m$. Away from the pole, the self-energy $\Sigma(q)$ is gauge-dependent, and any mass function extracted from it changes with $\xi$. The mass has thus been an \emph{on-shell concept}, with no gauge-invariant meaning at arbitrary virtuality.

The present construction overcomes this. The gauge-invariant self-energy $\hat\Sigma^{\text{ren}}(q)$, derived from the segment-local WTI at off-shell internal momentum and renormalized by on-shell subtraction, defines $m(q) = m + \hat\Sigma^{\text{ren}}(q)$ at every $q$. The concept of fermion mass thereby {\em extends} from a single number to a function of the momentum.

This extension is not merely formal but is forced by the structure of the gauge theory itself, in that the segment-local WTI generates gauge-invariant boundary terms at each fermion endpoint, and these define the dressed propagator $\hat\Sf^{\text{ren}}(q) = i/(\slashed{q}-m(q))$ at every momentum. The fermion mass is therefore not a single number but a gauge-invariant, Dirac-valued function of the momentum, a consequence of the Ward--Takahashi identity acting on the unamputated amplitude rather than a choice of scheme or truncation. The physical mass $m$ is only the special value $m(m)=m$; the value $m(q)$ at $\slashed{q}\neq m$ is equally physical and equally gauge-invariant, so the on-shell mass is one value of the mass function rather than a privileged number.

The gauge-invariant mass function $m(q)$ has immediate applications. By Lorentz covariance its scalar part depends only on $q^2$, and in contexts where a scalar mass function is needed one can project $m(q)$ onto its scalar component $m(q^2)$.

In processes with far-off-shell virtual fermions, $m(q)$ enters naturally and its gauge invariance ensures unambiguous results where the conventional self-energy would not.

\subsection{Observability of $m(q^2)$}

Gauge invariance is necessary but not sufficient for observability. Physical observables are $S$-matrix elements, which involve the full amplitude, not individual propagators. Nevertheless, $m(q^2)$ is observable in the same sense that the running coupling $\alpha(q^2)$~\cite{Grunberg:1984,Watson:1997} is, in that it is not measured directly but can be unambiguously extracted from cross sections at different kinematics, yielding the same result regardless of the process used. The PT decomposition that defines $\hat\Sigma$ is the unique decomposition preserving the Ward identity structure of the theory, so the extraction is not an arbitrary rearrangement but is dictated by the gauge symmetry itself.

The comparison with the running coupling makes the role of the renormalization scale explicit. Any scheme expresses a prediction through a scale-dependent coupling $\alpha(\mu)$ and a scale-dependent mass parameter, but the scale $\mu$ is not an observable and its dependence cancels in every $S$-matrix element. What survives for the coupling is the physical running coupling $\alpha(q^2)$ at the momentum the process probes, and what survives for the mass is the gauge-invariant function $m(q)$. The directly measured mass is the on-shell value, the pole of the propagator where $p^2 = m^2(q)$, which is the physical mass, while the off-shell values of $m(q)$ enter amplitudes at the corresponding momenta exactly as $\alpha(q^2)$ does. The one place the analogy is not exact is that $\alpha(q^2)$ is renormalization-group invariant and is therefore literally the same function in every scheme, whereas $m(q^2)$ is fixed by the on-shell conditions and is unique in that sense rather than scheme-invariant as a function. The scalar mass function $\hat M(q^2)$ of Section~\ref{sec:IR} closes even that gap: being multiplicatively renormalization-group invariant, it is the same function in every scheme once the pole mass is fixed, and it stands to the scheme parameter $\bar m(\mu)$ exactly as $\alpha(q^2)$ stands to the scheme coupling. What is convention-free on the mass side is the gauge-invariance of the whole function together with its pole value, and it is this gauge-invariance, absent from the conventional off-shell self-energy, that the present construction supplies.

The observability of the momentum-dependent mass has been discussed in detail in~\cite{Choi:2023:observables,Choi:2024cbs,Choi:2024hkd,Choi:2025:Higgs}, where it is shown that the renormalized mass, expressed as a difference of the self-energy at different scales, is finite, independent of the regularization scheme and insensitive to ultraviolet physics. These properties follow from the same structure that appears here, where on-shell subtraction removes the regularization-dependent pieces and what remains is a fully determined function of the momentum.

Since the on-shell scheme fixes all parameters in terms of physical observables ($m_{\text{pole}}$ and $e$), $m(q^2)$ is a unique, fully determined function with no residual scheme dependence. The off-shell mass function $m(q^2)$, expressed in terms of physical (on-shell) parameters, is (i)~gauge-invariant ($\partial m/\partial\xi=0$ for all $q^2$), (ii)~process-independent (by the locality of the segment of the WTI), (iii)~scheme-independent (one-to-one correspondence between renormalization schemes; conversion to on-shell parameters gives a unique result) and (iv)~uniquely decomposed for the canonical reference $\xi_0=1$ (gauge independence selects $\alpha=1$ in the longitudinal sector, the transverse vertex is separately gauge-invariant and does not participate in the reshuffling, and the residual reference-gauge family of Section~\ref{sec:uniqueness} consists of physically equivalent dressings rather than a genuine ambiguity). These four properties make $m(q^2)$ a genuine physical function of the momentum rather than a scheme artifact, observable in the operational sense that it is what every prediction reduces to once the unphysical scale cancels, with its on-shell value the directly measured mass. Away from the pole it is defined at fixed infrared regulator, with the infrared-finite $\hat M(q^2)$ of Section~\ref{sec:IR} available when a regulator-free function is required.

\section{Non-Abelian Extension}
\label{sec:qcd}

We outline the non-abelian extension here. The complete one-loop result is the quark-line $C_F$ inheritance derived below, together with the explicit $n=1$ color decomposition of Section~\ref{sec:qcd_n2}; the genuinely non-abelian multi-segment generalization to arbitrary $n$, with external gluons and the full pinch-plus-ghost structure on gluon segments, is left as an open structural problem (Section~\ref{sec:qcd_general_n}).

The framework has been formulated for general linear covariant gauge theories, with the QCD case appearing as the $C_2 = C_F$ reading of every expression at the level of the quark line. In particular, the quark-line analysis of Section~\ref{sec:explicit} produces, with no further work, the identification
\begin{equation}
  \hat\Sigma^{\rm QCD}(q)\Big|_{\text{quark-line sector}}
  = C_F \hat\Sigma^{\rm QED}(q)\Big|_{e\to g},
  \label{eq:CFinheritance}
\end{equation}
because the fundamental identity \eqref{eq:fund_intro} is purely Dirac-algebraic, the color factor $t^at^a = C_F \mathbf{1}_F$ commutes through the propagators, and the segment-locality argument of Section~\ref{sec:segments} sees no color whenever the segment endpoints are quark-gauge-boson vertices. The four-sector decomposition \eqref{eq:four_sectors}, the fourteen-term pinched amplitudes, the Dirac-shift cancellation table of Section~\ref{sec:cancel} and the four boundary terms $(2,2)$, $(+,1)$, $(2,-)$, $(+,-)$ all carry over row by row with the substitution $e^2 \to g^2 C_F$ for the internal-gauge-boson factor; the cancellations of $(+,1)$, $(2,-)$ by external quark self-energies and of $(+,-)$ by the $u$-channel proceed identically.

The non-trivial content of the non-abelian extension lies elsewhere. Three new structures appear, each associated with a different feature absent from QED. The three-gluon self-coupling introduces an algebraic decomposition that plays the role of the WTI on gluon segments, the ghost sector provides the diagrammatic completion of the cancellation and the color algebra adds bookkeeping that enlarges the cancellation table without disrupting its structural form. These three pieces are the content of the present section. The two genuinely new vertices needed throughout are the three-gluon vertex
\begin{equation}
  \Gamma^{abc}_{\alpha\mu\nu}(q,k_1,k_2) = gf^{abc} 
    \bigl[g_{\alpha\mu}(q-k_1)_\nu
        + g_{\mu\nu}(k_1-k_2)_\alpha
        + g_{\nu\alpha}(k_2-q)_\mu\bigr],
  \label{eq:3g}
\end{equation}
(all momenta incoming, $q+k_1+k_2=0$), and the ghost-gluon vertex
\begin{equation}
  V^{abc}_\mu(p,k) = -gf^{abc} p_\mu,
\end{equation}
where $p$ is the outgoing-ghost momentum. The four-gluon vertex contributes only at higher orders and does not enter the one-loop quark self-energy with external photons.

\subsection{Three-gluon vertex: pinch decomposition}
\label{sec:cornwall}

The first new ingredient appears whenever an internal gluon segment contains a three-gluon vertex. At one loop with external photons such a configuration does not arise for the quark self-energy, because the external lines are color-neutral and the loop carries a single gluon line that connects directly to the quark. The three-gluon vertex enters naturally at two loops, where the internal gluon develops a one-loop self-energy with two trilinear vertices, or at one loop in any process with external gluons.

The structural device that replaces the WTI on gluon segments is the pinch decomposition~\cite{Cornwall:1982,Cornwall:Papavassiliou:1989} of the three-gluon vertex into a piece with abelian Ward-identity structure and a residual pinching piece
\begin{equation}
  \Gamma^{abc}_{\alpha\mu\nu}(q,k_1,k_2)
  = \Gf{}^{abc}_{\alpha\mu\nu}(q,k_1,k_2)
  + \Gp{}^{abc}_{\alpha\mu\nu}(q,k_1,k_2),
  \label{eq:cornwall}
\end{equation}
with the explicit forms
\begin{equation}
\begin{split}
  \Gf{}^{abc}_{\alpha\mu\nu}(q,k_1,k_2) &= gf^{abc}\bigl[(k_1-k_2)_\alpha\, g_{\mu\nu} + 2q_\nu\, g_{\alpha\mu} - 2q_\mu\, g_{\alpha\nu}\bigr],\\
  \Gp{}^{abc}_{\alpha\mu\nu}(q,k_1,k_2) &= gf^{abc}\bigl[k_{2\nu}\, g_{\alpha\mu} - k_{1\mu}\, g_{\alpha\nu}\bigr],
\end{split}
  \label{eq:cornwall_explicit}
\end{equation}
with all momenta incoming as in~\eqref{eq:3g}. By construction $\Gf$ satisfies the abelian-like Ward identity
\begin{equation}
  q^\alpha \Gf{}^{abc}_{\alpha\mu\nu}(q,k_1,k_2)
  = -i\,gf^{abc}\bigl[D^{-1}_{\mu\nu}(k_2) - D^{-1}_{\mu\nu}(k_1)\bigr]_{\rm tree},
  \label{eq:GammaF_WTI}
\end{equation}
the $q_\mu q_\nu$ terms cancelling so that only the $g_{\mu\nu}$ inverse-propagator difference survives, while $\Gp$ collects the remaining terms. The two-term difference is in fact a property of the full vertex, whose contraction with the pinch momentum reads
\begin{equation}
  q^\alpha \Gamma^{abc}_{\alpha\mu\nu}(q,k_1,k_2)
  = gf^{abc}\bigl[(k_2^2\, g_{\mu\nu} - k_{2\mu}k_{2\nu})
  - (k_1^2\, g_{\mu\nu} - k_{1\mu}k_{1\nu})\bigr],
  \label{eq:full3g_STI}
\end{equation}
a difference of transverse tensors, of which $\Gf$ carries the $g_{\mu\nu}$ part \eqref{eq:GammaF_WTI} and $\Gp$ the longitudinal remainder $k_{1\mu}k_{1\nu}-k_{2\mu}k_{2\nu}$, whose adjacent-line momenta fire the secondary pinches. The right-hand side of \eqref{eq:GammaF_WTI} is the structural analog of the QED relation $\slashed\ell = i\bigl[\SfI(k+\ell)-\SfI(k)\bigr]$, in that when contracted against adjacent gluon propagators, $\Gf$ produces a difference of inverse propagators that telescopes within its segment exactly as the WTI does on quark segments.

The piece $\Gp$, by contrast, generates pinches that displace gauge-invariance-breaking contributions to neighboring topologies, just as the longitudinal photon piece $\Dp$ does in QED. Concretely, the relevant terms in $\Gp$ carry the longitudinal momenta $k_{1\mu}$ or $k_{2\nu}$, which upon contraction with the adjacent gluon propagators $D(k_1)$ or $D(k_2)$ pinch out the line and leave a tree-level structure to be reabsorbed. These reabsorbed pieces do not shift the intermediate-line propagator, so they never enter the gauge-invariant fermion self-energy $\hat\Sigma$; they are reassigned to the singlet and crossed color sectors, which build the modified three-gluon vertex $\hat\Gamma_{\rm 3g}$ and the $u$-channel amplitude, as worked out explicitly in Section~\ref{sec:qcd_n2}.

The upshot for the segment picture is that gluon segments admit the same local-firing logic as quark segments. A pinch insertion on a gluon segment, after the pinch decomposition, produces a difference of gluon propagators that telescopes within the segment, plus residual $\Gp$-contributions that propagate to adjacent topologies in a controlled way. The boundary-and-interior classification of Section~\ref{sec:segments} carries over with this enlarged notion of firing; in particular, the double-firing decomposition into the four sectors \eqref{eq:four_sectors} persists, with the two terms of the $\Gf$ difference supplying the two firing outcomes at a gluon endpoint (Section~\ref{sec:qcd_cancel}).

\subsection{Ghost contributions}
\label{sec:ghosts}

The pinch decomposition \eqref{eq:cornwall} alone does not produce a gauge-invariant gluon self-energy. The Ward identity \eqref{eq:GammaF_WTI} holds only for the tree-level inverse propagator; at one loop the correct identity that $\Gf$ satisfies is the Slavnov--Taylor identity, which involves ghost Green's functions on the right-hand side. The ghost loops in the conventional Feynman-gauge gluon self-energy supply precisely these missing terms.

Diagrammatically, the role of ghosts is to render the kinematic combination
\begin{equation}
  \Pi^{ab}_{\mu\nu}(q)\Big|_{\Gf} + \Pi^{ab}_{\mu\nu}(q)\Big|_{\rm ghost}
\end{equation}
transverse and gauge-invariant, whereas neither term is transverse on its own. In our segment language, the ghost loop is best understood as a ``virtual segment'' running in parallel with each gluon segment whose endpoints carry three-gluon vertices, and whose contribution must be added before the cancellation table closes.

For the purposes of bookkeeping, every gluon segment that carries internal trilinear gauge structure brings a paired ghost contribution into the cancellation table. In contrast to QED, where ghost rows simply do not exist, in QCD the cancellation table acquires ghost rows that must be matched against the corresponding gluon-loop and pinch-vertex contributions.

\subsection{External gluons: new topologies}
\label{sec:external_gluons}

A second route to non-abelian content is to replace the external test photons of the Compton process by external gluons, so that the embedding amplitude is the off-shell quark-gluon scattering amplitude $qg_1 \to qg_2$. This setup is closer to physical QCD applications and exposes the new vertex structure already at tree level.

At tree level, in addition to the QED-like $s$- and $u$-channel quark exchange diagrams, a third tree topology exists in which the two external gluons join at a three-gluon vertex and a single internal gluon line connects to the quark
\begin{equation}
  \mathcal{T}_{3g}
  = \bar u(p') (igt^c\gamma^\rho) u(p)
    D_{\rho\alpha}(k_1+k_2)
    \Gamma^{abc}_{\alpha\mu\nu}(k_1+k_2,-k_1,-k_2)
     \epsilon_1^{a \mu} \epsilon_2^{*b \nu}.
\end{equation}
At one loop, the topologies of Section~\ref{sec:explicit} multiply because each of the four configurations $\mathcal{S}$, $\mathcal{R}$, $\mathcal{L}$, $\mathcal{B}$ admits the additional possibility that the internal gluon line attaches to one of the external gluon lines via a three-gluon vertex, and analogous ghost diagrams must be included.

The segment-locality framework now distinguishes two types of segments. {\em Quark segments} are bounded by quark-gluon vertices and admit the WTI \eqref{eq:fund_intro} as their firing rule, just as in the QED case, with color generators commuting through the Dirac structure trivially. {\em Gluon segments} are bounded by three-gluon vertices or by external gluon legs, and on these the firing mechanism is the pinch decomposition \eqref{eq:cornwall} together with paired ghost contributions. The boundary of a gluon segment is the three-gluon vertex at its endpoint, which is no longer inert but carries non-trivial momentum dependence and contributes to the firing.

A pinch insertion now telescopes within whichever type of segment it sits on, with the appropriate firing rule, and the boundary-and-interior classification of Section~\ref{sec:segments} applies to each type separately.

\subsection{Generalized cancellation pattern}
\label{sec:qcd_cancel}

The organizing structure of the abelian cancellation survives intact. The pinch gluon still has two endpoints, each firing still produces a signed two-term difference, and the pinched amplitude still decomposes into the four sectors of \eqref{eq:four_sectors},
\begin{equation}
  \mathcal{M}^P = \mathcal{M}_{LL} + \mathcal{M}_{LR} + \mathcal{M}_{RL} + \mathcal{M}_{RR},
\end{equation}
with the universal $(+,-,-,+)$ sign pattern. Only the firing rule at an endpoint changes with the segment type. At a quark endpoint the two terms are the inverse quark propagators of the WTI \eqref{eq:fund_intro}, exactly as in QED; at a gluon endpoint they are the inverse gluon propagators of the $\Gf$ Ward identity \eqref{eq:GammaF_WTI}, which assigns the roles of $L$ and $R$ to the two adjacent gluon lines. Explicitly, for a pinch insertion splitting a gluon segment into lines of momenta $k$ and $k+\ell$, the $\Gf$ firing telescopes as
\begin{equation}
  D(k)\,\bigl[\ell^\alpha\Gf{}_\alpha\bigr]\,D(k+\ell)
  = -i\,gf^{abc}\bigl[D(k) - D(k+\ell)\bigr]
  \label{eq:gluon_telescope}
\end{equation}
(Lorentz indices suppressed), the same two-term difference with the same relative minus sign as the quark-segment identity \eqref{eq:fund_intro}; the color factor and the overall $-i$ are common to all four sectors of a placement and do not disturb the $(+,-,-,+)$ pattern. One qualification distinguishes the gluon case: quark propagators are gauge-independent, so the quark-segment telescoping is exact in any $\xi$, whereas gluon segment propagators are themselves $\xi$-dependent, and \eqref{eq:gluon_telescope} holds exactly for their $g_{\mu\nu}$ part, in particular in Feynman gauge, while their $(1-\xi)$ longitudinal parts generate secondary pinches. These secondary pinches, together with the $\Gp$ remainder and the paired ghost contributions, fall outside the four sectors and enter the table as additional rows, described below. The $(a,b)$ bookkeeping of Section~\ref{sec:general_n} therefore applies verbatim to the WTI-plus-$\Gf$ part of every diagram, on quark and gluon segments alike, and the three modifications that follow concern the color weights and the extra rows, not the sector structure.

With this understood, the Dirac-shift cancellation table of Section~\ref{sec:cancel} extends to QCD with three modifications.

First, every entry acquires a color factor that depends on the topology and on the color structure of the contributing diagram. For diagrams in the quark-line sector [eq.~\eqref{eq:CFinheritance}], the factor is uniformly $C_F$ and the classification is unchanged in structure. For diagrams involving three-gluon vertices, the color factor involves $f^{abc}$ contractions that, when combined with the $t^a$ generators on the quark line, can be simplified using
\begin{equation}
  if^{abc}t^bt^c = -\tfrac12 C_A t^a, \qquad C_A = N,
\end{equation}
producing entries whose net color weight is $C_A C_F$ or, after combination with the box-like topologies, $(C_F - \tfrac12 C_A)$.

Second, the classification acquires new entries corresponding to Dirac structures that are accompanied by gluon momentum shifts on internal gluon lines. These entries have no QED analog; they exist because, on a gluon segment, the pinch decomposition \eqref{eq:cornwall} can shift gluon propagator momenta in the same way that the WTI shifts quark propagator momenta on a quark segment. In a notation parallel to Section~\ref{sec:cancel}, these entries are identified by naming the additional gluon lines that carry a shift.

Third, the classification acquires ghost entries. A ghost loop is structurally a closed scalar loop with two or more ghost-gluon vertices, and it contributes to the same Dirac structures as the corresponding gluon-loop diagrams it accompanies. The role of these entries in the cancellation is to render the $\Gf$ piece compatible with full gauge invariance, as discussed in Section~\ref{sec:ghosts}.

The interior-versus-boundary distinction is preserved by all three modifications. Interior entries still cancel pairwise, with cancellations now occurring among diagrams that include three-gluon-vertex contributions and ghost loops alongside the abelian-like rows. Boundary rows survive and, as in the abelian case, are cancelled either by external self-energy diagrams (for boundary terms with one shifted leg) or by crossed-channel diagrams (for boundary terms with all legs shifted).

\subsection{Multi-segment generalization}
\label{sec:qcd_general_n}

The $k$-telescoping argument of Section~\ref{sec:general_n} extends to QCD, but the two cases stand on different footing, and we separate them at the outset. With external photons on the quark line the generalization is a theorem: it reduces to the abelian result times $C_F$. With external gluons it is so far only a mechanism: the segment-local firing rules are in hand, but the proof that the cancellation table closes at arbitrary $n$ is not, and we state below precisely what remains open. Both rest on taking the segment classification of Section~\ref{sec:external_gluons} seriously.

For a quark line dressed with $n+1$ external photon vertices and one internal gluon, the analysis is the abelian one of Section~\ref{sec:general_n} multiplied by $C_F$, with the quark-line segment structure unchanged, the WTI firing segment-locally, and the $(a,b)$ sum rule closing exactly as in QED. This generalizes the quark-line-sector inheritance~\eqref{eq:CFinheritance} from $n=1$ to arbitrary $n$.

For a quark line dressed with $n+1$ external gluon vertices, the $(a,b)$ analysis must be supplemented by a parallel structure on gluon segments. Each external gluon vertex now bounds a gluon segment as well as a quark segment, and pinch insertions can land on either. The two segment types interleave but do not interact at the level of the local firing rule. Quark-segment firings continue to follow the abelian WTI, gluon-segment firings follow the pinch decomposition with ghost completion, and the interior $(a,b)$ cancellation argument applies to the full set of quark-and-gluon shift patterns.

The number of contributing topologies grows accordingly. For external photons the count at general $n$ is the same as in the abelian case ($n+2$ topologies for the open-line analysis). For external gluons the count grows because each external vertex can be a color-singlet attachment or can sit on an internal three-gluon-vertex web. A complete enumeration of these topologies in our segment language and a proof that the cancellation table closes for arbitrary $n$ is the natural non-abelian counterpart of the theorem of Section~\ref{sec:general_n}, and is the main structural problem opened by the present construction.

This open item concerns the segment-local re-derivation in the present language, not the gauge invariance of the self-energy itself. The gauge-invariant QCD quark self-energy, with its three-gluon-vertex and ghost content, is an established result of the pinch technique, constructed explicitly at one and two loops and to all orders through the background-field correspondence~\cite{Cornwall:1982,Cornwall:Papavassiliou:1989,Binosi:Papavassiliou:2002,Binosi:Papavassiliou:2004,Binosi:Papavassiliou:2009}. The mass function needs only this two-point function, which the $n=1$ embedding of Section~\ref{sec:qcd_n2} already reproduces as $C_F$ times the Feynman-gauge QED result~\eqref{eq:hatSigma_QCD}, so it rests on the same footing in QCD as in QED. The arbitrary-$n$ external-gluon analysis would add the segment-local, amplitude-embedded form of that statement, not the statement itself.

In summary, the pinch-technique calculation of the off-shell quark self-energy in QCD splits into three structural components, of which the first is strictly abelian, the second introduces the genuine non-abelian content and the third is the diagrammatic completion required for gauge invariance.

The quark-line sector is built from the QED topologies with $t^at^a=C_F$ and is unchanged from Section~\ref{sec:explicit}, carrying an overall factor $C_F$. The three-gluon-vertex sector, organized by the pinch decomposition $\Gamma=\Gf+\Gp$, supplies the WTI analog on the gluon segments. The ghost sector, made of Faddeev--Popov loops, completes the gauge invariance of the $\Gf$ contribution.

The off-shell external-quark and external-gluon configuration, with boundary terms tracked diagrammatically, is the natural framework for the multi-segment generalization of Section~\ref{sec:general_n} to non-abelian theories. The abelian case extends straight through with a $C_F$ multiplier; the genuinely non-abelian extension, with external gluons and the full pinch plus ghost structure on gluon segments, is to our knowledge not yet treated in this language in the literature, and represents the natural continuation of the present construction.

\section{Example: Off-Shell Quark-Gluon Compton Amplitude}
\label{sec:qcd_n2}

We now apply the framework of the previous section to the minimal non-abelian embedding for $\hat\Sigma$, the off-shell quark-gluon $s$-channel amplitude. With two external gluons flanking an internal quark propagator at momentum $q$, the framework extracts $\hat\Sigma^{\rm QCD}(q)$ explicitly and exhibits how it reduces to a $C_F$-multiple of the abelian result while the genuinely non-abelian content cancels into other sectors.

The tree amplitude has the $s$-channel structure
\begin{equation}
  \mathcal{M}_0^{a_1 a_2,\mu\nu}
  = \Sf(p') \bigl(igt^{a_2}\gamma^\nu\bigr) 
    \Sf(q) \bigl(igt^{a_1}\gamma^\mu\bigr) \Sf(p),
\end{equation}
with off-shell external quarks $p,p'$ and external gluons $(k_1,a_1,\mu)$ and $(k_2,a_2,\nu)$, related by $q = p+k_1 = p'-k_2$. 

\subsection{Diagram inventory at one loop}

Three classes of one-loop diagrams contribute gauge-dependent pieces through the longitudinal gluon.

(i) Abelian-like topologies. The internal gluon connects two points on the quark line via quark-gluon vertices, exactly as the QED loop photon did in Section~\ref{sec:explicit}. There are four such topologies $\mathcal{S}$, $\mathcal{R}$, $\mathcal{L}$, $\mathcal{B}$ with the same geometric assignment as in the abelian case, $\mathcal{S}$ having both endpoints on the $q$-segment, $\mathcal{R}$ one on the $q$-segment and one on the $p$-segment, $\mathcal{L}$ one on the $q$-segment and one on the $p'$-segment and $\mathcal{B}$ one endpoint on each external leg.

(ii) Non-abelian topologies. The internal gluon attaches via a three-gluon vertex to one of the external gluons, with its other endpoint on the quark line. There are six such diagrams, indexed by which external gluon hosts the three-gluon vertex (two choices) and which quark-line segment carries the other endpoint (three choices).

(iii) External quark self-energies on the $p$-leg and $p'$-leg, identical in structure to the abelian case.

Ghost diagrams do not enter at one loop in the $s$-channel quark exchange. They appear at two loops through the gluon self-energy that dresses the internal line, and at one loop only if the external gluon lines are explicitly dressed. Neither configuration affects the analysis here.

\subsection{Color decomposition}

The color factors of the abelian-like topologies are
\begin{align}
  \mathcal{S}: &\quad t^bt^b \text{ inside }q\text{-segment}
    \Rightarrow C_F t^{a_2}t^{a_1}, \\[3pt]
  \mathcal{R},\mathcal{L}: &\quad t^bt^{a_i}t^b\text{ flanking one external vertex}
    \Rightarrow \bigl(C_F - \tfrac12 C_A\bigr) t^{a_2}t^{a_1}, \\[3pt]
  \mathcal{B}: &\quad t^b t^{a_2}t^{a_1} t^b
    \Rightarrow \tfrac{1}{4}\delta^{a_2 a_1}\mathbf{1}_F
       + \bigl(C_F - \tfrac12 C_A\bigr) t^{a_2}t^{a_1}.
\end{align}
The second line uses $t^bt^at^b = (C_F-\tfrac12 C_A)t^a$. The third uses the SU($N$) Fierz identity $t^b_{ij}t^b_{kl} = \tfrac12\bigl(\delta_{il}\delta_{jk}-\tfrac{1}{N}\delta_{ij}\delta_{kl}\bigr)$ applied to the four-generator sandwich, with $\mathrm{Tr}(t^{a_2}t^{a_1})=\tfrac12\delta^{a_2 a_1}$ and $-\tfrac{1}{2N}=C_F-\tfrac12 C_A$.

The non-abelian topologies carry $f^{abc}t^bt^c=\tfrac{i}{2}C_A t^a$ on the quark-line factor (with sign depending on the orientation of the three-gluon vertex), giving color factors of the form $C_A t^{a_2}t^{a_1}$, $C_A t^{a_1}t^{a_2}$ or $C_A \delta^{a_2 a_1}\mathbf{1}_F$ after the three-gluon vertex tensor is contracted out.

External quark SEs have color $C_F t^{a_2}t^{a_1}$, since the internal SE loop closes on $t^bt^b=C_F$ and the external vertices $t^{a_1}$ and $t^{a_2}$ sit outside this loop and commute through.

The contributions organize into three independent color sectors, indexed by the color tensor they multiply
\begin{itemize}
\item The {\em $t^{a_2}t^{a_1}$ sector}, the tree-level color structure. Both $C_F$ and $C_A$ pieces of the abelian-like topologies land here, together with the $t^{a_2}t^{a_1}$ part of the non-abelian topologies and the external SEs.
\item The {\em $t^{a_1}t^{a_2}$ sector}, generated only by non-abelian topologies.
\item The {\em $\delta^{a_2 a_1}\mathbf{1}_F$ sector}, generated by the singlet part of topology $\mathcal{B}$ together with matching non-abelian topologies.
\end{itemize}
The three sectors close independently. The gauge-invariant fermion self-energy $\hat\Sigma(q)$ lives in the $t^{a_2}t^{a_1}$ sector, and specifically in its $C_F$ projection. The remaining content of the $t^{a_2}t^{a_1}$ sector (its $C_A$ piece), together with the entire $t^{a_1}t^{a_2}$ and $\delta^{a_2 a_1}$ sectors, feeds other gauge-invariant Green's functions, principally the modified three-gluon vertex $\hat\Gamma_{\rm 3g}$, not $\hat\Sigma$.

\subsection{The $C_F$ projection: QED cancellation embedded}

Restricting to the $C_F$ projection of the $t^{a_2}t^{a_1}$ sector, the $s$-channel topology $\mathcal{S}$ and the external self-energy pinches on the $p$- and $p'$-legs carry the full $C_F$, while $\mathcal{R}$, $\mathcal{L}$ and $\mathcal{B}$ each contribute the $C_F$ piece of their $C_F-\tfrac12 C_A$ color factor. The non-abelian topologies have no $C_F$ part and survive only at order $C_A$.
With color factored out, the Dirac structure of each topology is identical to its QED counterpart, because the WTI firing depends only on Dirac structure and the spectator color generators commute through. The fourteen-term decomposition of Section~\ref{sec:cancel} reproduces row by row
\begin{equation}
  X^{\rm QCD}\big|_{C_F}
  = C_F t^{a_2}t^{a_1} X^{\rm QED},
  \qquad X \in \{\mathcal{S},\mathcal{R},\mathcal{L},\mathcal{B}\},
\end{equation}
and the external SE pinches likewise inherit a $C_F$ multiplier.

The three interior cancellations \eqref{eq:cancel_010}, \eqref{eq:cancel_011}, \eqref{eq:cancel_110} therefore close identically. The four surviving boundary terms of \eqref{eq:surviving}, namely $(a,a)$, $(+,b)$, $(a,-)$ and $(+,-)$, appear with the same coefficients and the same shift structure, in that only $(+,-)$ has the intermediate-line propagator shifted, so all of $(a,a)$, $(+,b)$ and $(a,-)$ again leave $\Sf(q)$ untouched, confirming that the gauge-invariant intermediate-line self-energy is unmodified by boundary effects. The $(a,a)$ term is removed by on-shell renormalization as in Section~\ref{sec:renorm_mass}. The $(+,b)$ and $(a,-)$ terms are cancelled by the external SE pinches on the $p$- and $p'$-legs, whose color $C_F t^{a_2}t^{a_1}$ matches exactly. The $(+,-)$ term is cancelled by the $u$-channel of the QCD Compton-like amplitude, with the crossed color structure $t^{a_1}t^{a_2}$ projecting onto its own row in the cancellation table.

The conclusion of Section~\ref{sec:explicit} carries over verbatim to the $C_F$ projection
\begin{equation}
  \hat\Sigma^{\rm QCD}(q)
  = C_F \hat\Sigma^{\rm QED}(q)\Big|_{e\to g}
  = C_F \Sigma_\xi^{\rm QED}(q)\Big|_{\xi=1, e\to g}.
  \label{eq:hatSigma_QCD}
\end{equation}
This reproduces the standard one-loop result for the gauge-invariant quark self-energy in QCD~\cite{Cornwall:1982,Binosi:Papavassiliou:2002}, derived here without invoking the Dirac equation or any on-shell projection on the external quarks or gluons.

\subsection{The $C_A$ projection of the $t^{a_2}t^{a_1}$ sector}

The $C_A$ pieces of $\mathcal{R}$, $\mathcal{L}$, $\mathcal{B}$ (color $-\tfrac12 C_A t^{a_2}t^{a_1}$) populate the same Dirac shift rows of Section~\ref{sec:cancel} as their $C_F$ counterparts, but with sign reversed. The non-abelian topologies, after the pinch decomposition of each three-gluon vertex, contribute terms with color $+\tfrac12 C_A t^{a_2}t^{a_1}$ to the same Dirac rows.

The mechanism is the following. The $\Gf$ piece of the pinch decomposition \eqref{eq:cornwall}, when contracted with the longitudinal $\ell$ from the internal gluon, produces a difference of inverse gluon propagators via \eqref{eq:GammaF_WTI}. These inverse propagators eat the adjacent gluon propagators, and the non-abelian topology reduces to a quark-line factor with the same Dirac structure as the corresponding abelian-like topology but with color $\tfrac12 C_A t^{a_2}t^{a_1}$. Row by row, the $-\tfrac12 C_A$ contribution from the abelian-like topology is cancelled by the $+\tfrac12 C_A$ contribution from the pinch-reduced non-abelian topology, and the $C_A$ projection of the $t^{a_2}t^{a_1}$ sector closes with the same three interior cancellations as the $C_F$ projection.

The boundary terms $(a,a)$, $(+,b)$, $(a,-)$, $(+,-)$ survive in the $C_A$ projection as well, but here they have a different fate. The $(a,a)$ term is again $q$-independent and removed by the on-shell subtraction in the renormalized mass. The $(+,b)$ and $(a,-)$ terms are not cancelled by external quark SEs (which are pure $C_F$) but by the $\Gf$-induced pinches on the non-abelian topologies attached at the external gluon vertices. These supply the non-abelian wave function correction to the external gluon line. The $(+,-)$ term is cancelled by the $u$-channel of the non-abelian topologies in the same way as the abelian-like $(+,-)$. None of the $C_A$ projection boundary cancellations shift the intermediate-line propagator $\Sf(q)$, so none of them contribute to $\hat\Sigma$. They define instead a gauge-invariant correction to the external gluon-leg dressing, which is absorbed into the modified vertex $\hat\Gamma_{\rm 3g}$ for the external gluon.

\subsection{The new color sectors}

The $t^{a_1}t^{a_2}$ sector is populated by the non-abelian topologies with the three-gluon vertex on the opposite external gluon to the abelian-like configuration. Its cancellation runs in parallel to the $t^{a_2}t^{a_1}$ sector, with the same pinch mechanism producing interior cancellations and surviving boundary terms. The Dirac structure of the surviving boundary terms is proportional to the $u$-channel tree amplitude. These contributions define the $u$-channel gauge-invariant amplitude and do not enter $\hat\Sigma$.

The $\delta^{a_2 a_1}\mathbf{1}_F$ singlet sector receives its non-cancelling abelian-like contribution from the singlet piece of topology $\mathcal{B}$, and its non-abelian-topology partner from the singlet projection of the $\Gp$ pieces. This sector contributes to the modified three-gluon vertex $\hat\Gamma^{abc}_{\rm 3g}$ at the external gluon attachments, which is a separate gauge-invariant Green's function with its own pinch identification (the gluon self-energy via PT in the context where it is the primary object). The structural analysis is the same as in the $t^{a_2}t^{a_1}$ sector but with the singlet color tensor playing the role of the tree-level color structure.

\subsection{Summary at $n=1$}

The $n=1$ QCD Compton-like embedding closes the cancellation table in every color sector. The gauge-invariant fermion self-energy $\hat\Sigma^{\rm QCD}(q)$ lives entirely in the $C_F$ projection of the tree color sector and equals $C_F$ times the QED result in Feynman gauge, eq.~\eqref{eq:hatSigma_QCD}. The $C_A$ projection of the same color sector closes against the pinch-reduced non-abelian topologies and contributes to the modified gluon-vertex sector, not to $\hat\Sigma$. The new color sectors $t^{a_1}t^{a_2}$ and $\delta^{a_2 a_1}$ are populated entirely by non-abelian content and contribute to the $u$-channel and to the three-gluon vertex, respectively.

Segment locality on the quark line operates throughout. The WTI fires within each quark segment regardless of which color sector the diagram belongs to, with color generators commuting through Dirac propagators as spectators. The non-abelian content of the theory appears at exactly two places, the color reduction step that distributes each topology among the color sectors and the diagrammatic substitution of the three-gluon vertex by its pinch decomposition on gluon segments. Neither of these disrupts the segment-local firing of the WTI on quark segments.

The $n=1$ QCD Compton-like amplitude is therefore the minimal embedding that delivers $\hat\Sigma^{\rm QCD}(q)$ as a closed and explicitly gauge-invariant object. The framework extends to general $n$ by combining the abelian $k$-telescoping of Section~\ref{sec:general_n} on the quark line with parallel pinch-mediated telescoping on gluon segments, the structural problem posed in Section~\ref{sec:qcd_general_n}.

\section{Extension to All Orders}
\label{sec:allorders}

The framework developed above (segment-local WTI, interior and boundary decomposition, generalized self-energy, off-shell mass) carries to all orders in perturbation theory with its conceptual structure intact, the exact Ward--Takahashi identity taking the place of its tree-level form. The argument is an induction on the loop order rather than a term-by-term recomputation, checked explicitly against the two-loop self-energy below. The non-abelian extension of Sections~\ref{sec:qcd}--\ref{sec:qcd_n2} is an independent generalization in a different direction. The two extensions are structurally orthogonal, with the all-orders argument operating within a fixed gauge group and the non-abelian argument at fixed loop order. Combining them to higher orders in QCD requires both pieces of machinery together and the Batalin--Vilkovisky formalism, as discussed at the end of this section.

\subsection{Segment locality and the reduction to linear chains}

This extension rests on the single statement of segment locality introduced in Section~\ref{sec:segments}, that the gauge-dependent part of a self-energy on an interval is cancelled by the local firing rule at the two vertices bounding that interval, and by nothing else, however complicated the surrounding diagram. The firing rule is the WTI on a quark segment and the pinch decomposition on a gluon segment. The remainder of this section establishes why this holds for an arbitrary diagram and how the cancellation propagates once it fires.

Charge conservation makes every fermion line a single unbranched chain, either an open line between two external legs or a closed loop, so the fermion skeleton of any diagram, with arbitrary loops and external legs, is a set of linear chains and cycles, and a self-energy occupies one interval of one such chain. Boson lines only attach points on these chains and never create a fermion branch, so there is a unique fermion path through a given interval, namely the chain that carries it.

Cancelling the gauge-dependent part of a self-energy on one interval requires more than firing the WTI at the two vertices bounding the interval, since each firing leaves a boundary term on a neighboring propagator. That boundary term is cancelled by the matching term from the adjacent placement on the same chain, which in turn passes a boundary term to its own neighbor, so the cancellation walks the chain and closes only at the chain's topological boundary. For an open chain that boundary is the external legs, where the residual terms are removed by the external-leg self-energies and the crossed channels, and for a closed loop there is no boundary and nothing survives. This walk is the $(a,b)$ sum rule of Section~\ref{sec:general_n}, generalizing at $L$ loops to the $L$-dimensional placement lattice below. The interval one started from is never shifted, so the self-energy on it is gauge-invariant, which is what defines the mass function.

In an abelian theory this reduction is complete, since the only structured lines are the matter chains and the photon, lacking a self-coupling, cannot branch. In a non-abelian theory the quark chains are unchanged and the quark-segment cancellation is still the linear-chain telescoping, but the gluon carries the three- and four-gluon vertices, so the gauge-boson sector branches into a web rather than a set of chains. Each gluon branch point is where the chain rule gives way to the pinch firing rule of Section~\ref{sec:qcd}, in that the $\Gf$ Ward identity~\eqref{eq:GammaF_WTI} is a telescoping step carried across a branch rather than along a chain, completed by the ghost terms. The gauge-sector branching is the orthogonal non-abelian ingredient and leaves the matter-chain reduction intact.

\subsection{The exact Ward--Takahashi identity}

The all-orders WTI in QED is
\begin{equation}
  \ell_\mu \Gamma^\mu(k',k) = i\bigl[\SfI(k')-\SfI(k)\bigr],
  \label{eq:exact_WTI}
\end{equation}
where $\Gamma^\mu(k',k)$ is the {\em full} (all-orders) proper vertex, normalized so that its tree-level value is $\gamma^\mu$ with the coupling factored out (with $k$ incoming, $k'=k+\ell$ outgoing), and $\SfI(k) = -i(\slashed k - m - \Sigma(k))$ is the {\em full} inverse propagator in the convention~\eqref{eq:PS_fermion} extended to all orders. At tree level \eqref{eq:exact_WTI} reduces to \eqref{eq:WTI_tree}. This identity is exact, following from $\partial_\mu J^\mu = 0$ in the path integral, with no perturbative truncation.

When sandwiched between full propagators, it gives the all-orders analog of the fundamental identity \eqref{eq:fund_intro}
\begin{equation}
  \Sf(k) \ell_\mu\Gamma^\mu(k',k) \Sf(k')
  = i\bigl[\Sf(k)-\Sf(k')\bigr],
  \label{eq:exact_fund}
\end{equation}
where now $\Sf(k)$ is the {\em full} dressed propagator. The cancellation $\Sf(k) \SfI(k)=\mathbf{1}$ is an algebraic identity that holds whether $\Sf$ is free or fully dressed.

\subsection{The segment-local mechanism is identical}

At any loop order, when $\ell^\mu$ from the longitudinal part $D^P_{\mu\nu}$ of any internal photon propagator hits a vertex, the exact WTI \eqref{eq:exact_WTI} fires at that vertex. The result is a difference of full inverse propagators, each of which eats its neighboring full propagator.

The only difference between one loop and all orders is the identity of the objects entering this move; the segment-local telescoping itself is unchanged. At one loop the Ward--Takahashi identity is the tree relation $\slashed\ell = i[\Sf^{(0)}]^{-1}(k+\ell)-i[\Sf^{(0)}]^{-1}(k)$, the propagator is the zeroth-order renormalized $\Sf^{(0)}(k)=i/(\slashed{k}-m)$ built from the physical mass at tree level, the vertex is the tree-level $\gamma^\mu$ and the cancellation driving the telescoping is the algebraic identity $\Sf^{(0)}[\Sf^{(0)}]^{-1}=\mathbf{1}$. At all orders each object is promoted to its dressed counterpart, so the identity becomes the exact $\ell_\mu\Gamma^\mu = e[\SfI(k')-\SfI(k)]$, the propagator the full $\Sf(k)=i/(\slashed{k}-m-\Sigma)$, the vertex the full proper vertex $\Gamma^\mu(k',k)$ and the cancellation $\Sf\,\SfI=\mathbf{1}$.

The telescoping proceeds segment by segment along the fermion line, exactly as at one loop. Interior terms cancel by the same segment-local mechanism. Boundary terms are generated at the fermion endpoints by the same segment-local mechanism. The boundary and interior decomposition is structural, not perturbative.

The content of the all-orders proof is fixed by a single requirement, that the entire $\xi$-dependent part of $\Sigma_\xi(q)$ be reconstructed by the pinch contributions order by order so that only $\Sigma_\xi(q)|_{\xi=1}$ remains. The gauge parameter enters solely through the longitudinal part $(1-\xi)\ell^\mu\ell^\nu/\ell^4$ of each internal photon propagator, and in QED this part is not renormalized, since the vacuum polarization is transverse and the longitudinal piece of the full photon propagator stays at its tree-level form at every order. The $\xi$-dependent part of $\Sigma_\xi$ is therefore a sum of terms in which some subset of the internal photons is taken longitudinal, one factor of $(1-\xi)$ for each, and each longitudinal photon is pinched on its own, its $\ell^\mu$ firing the exact WTI~\eqref{eq:exact_WTI} at the adjacent vertex. Because the identity is exact, this pinch is valid whatever the other photons are doing and whatever dressing the neighboring lines already carry, so the longitudinal photons are removed independently and a term with several of them longitudinal is the iterated single pinch organized by the lattice below.

Diagrammatically this is a recursive, self-similar structure that climbs the loop order by induction. Pinching a longitudinal photon at order $L$ produces the dressed difference $i[\SfI(k+\ell)-\SfI(k)] = \slashed\ell - [\Sigma(k+\ell)-\Sigma(k)]$, whose self-energy piece is of lower order, so the kinetic structure $(\slashed q - m)$ that made $\Sigma_P$ removable at one loop is preserved with $(\slashed q - m)$ replaced by the dressed analog $\hat\SfI(q)$, and the step consumes the gauge invariance already established through order $L-1$. Assuming gauge invariance through order $L-1$ and applying the exact WTI on the order-$(L-1)$ dressed objects therefore gives gauge invariance at order $L$ by the same segment-local move at every rung, with the exact WTI as the only ingredient absent at one loop. The decomposition~\eqref{eq:Ddecomp} applies recursively with dressed objects throughout~\cite{Choi:2025tus}. In a non-abelian theory the quark-segment step is identical, and the only addition is the gluon-segment firing rule of Section~\ref{sec:qcd}, activated once the internal gluon is dressed.

At $L$ loops there are $L$ internal photons, each with two endpoints distributed among the $n+1$ segments defined by the external photon vertices. Each pinch photon fires the WTI locally at each of its endpoints, producing a difference of dressed propagators with momenta differing by that photon's loop momentum. The one-loop $k$-telescoping pattern of Section~\ref{sec:general_n} generalizes to an $L$-dimensional lattice, the topological index $k$ becoming a vector $(k_1,\ldots,k_L)$, with telescoping cancellations along each direction independently and only the chain endpoints surviving. The within-segment structure becomes richer with each additional loop, but the segment boundaries remain inert because the external photon vertices are unchanged.

\subsection{All-orders self-energy and mass function}

The generalized self-energy at all orders is defined by the same prescription
\begin{equation}
  \hat\Sigma(q) = \Sigma_\xi(q)
  + \Delta\Sigma^{\text{pinch}}(q),
\end{equation}
where $\Sigma_\xi(q)$ is the conventional (all-orders) self-energy and $\Delta\Sigma^{\text{pinch}}(q)$ collects all propagator-like pinch contributions from vertex and box diagrams at every loop order, generated by the WTI acting on the internal fermion line. Since the pinch contributions at every order originate from $D^P_{\mu\nu} \propto(1-\xi)$ and cancel the gauge-dependent part of $\Sigma_\xi(q)$ with no remainder, the Feynman-gauge equivalence \eqref{eq:hatSigma_Feynman} extends to all orders
\begin{equation}
  \hat\Sigma(q) = \Sigma_\xi(q)\big|_{\xi=1}.
\end{equation}

The arguments for gauge invariance, process independence and the existence of $m(q)$ (Sections~\ref{sec:genSE}--\ref{sec:mass}) carry over to every order by the same induction, since the exact WTI guarantees that pinch contributions cancel the gauge dependence of $\Sigma_\xi(q)$ at every order and on-shell renormalization yields $m(q) = m + \hat\Sigma^{\text{ren}}(q)$ with $m(m)=m$ and unit residue. The two-loop result of Binosi and Papavassiliou~\cite{Binosi:Papavassiliou:2002}, $\hat\Sigma^{(2)} = \Sigma_\xi^{(2)}\big|_{\xi=1}$ in both QED and QCD, is the explicit verification at the smallest non-trivial loop order. We present the inductive step together with this two-loop anchor rather than a closed combinatorial proof at all orders; a complete proof would require controlling the $L$-dimensional placement lattice at every order.

\subsection{All orders in non-Abelian gauge theories}

The all-orders extension to QCD requires the apparatus of Sections~\ref{sec:qcd}--\ref{sec:qcd_n2} together with the all-orders machinery above. Three new structures enter, none of which violate segment locality.

First, the quark-segment WTI firing rule is replaced by the non-abelian WTI at the level of bare Green's functions and by the Slavnov--Taylor identity at the level of dressed ones, but the firing is still local to the quark segment and the color generators still commute through the Dirac propagators as spectators. The all-orders generalization of the Slavnov--Taylor identity is controlled by the Batalin--Vilkovisky (BV) formalism, developed for the pinch technique by Binosi and Papavassiliou~\cite{Binosi:Papavassiliou:2002:BV,Binosi:Papavassiliou:2004}. Within the BV framework, the segment-local firing rule on quark segments retains its abelian-WTI form at all orders, with bare propagators and vertices replaced by dressed ones, exactly as in the QED case; each double firing therefore still decomposes into the four sectors of \eqref{eq:four_sectors} with the $(+,-,-,+)$ pattern.

Second, gluon segments become active at two loops and beyond. At one loop in the $s$-channel quark exchange, the internal gluon is a single bare line and has no segment structure of its own. At two loops the internal gluon is dressed by its one-loop self-energy, which contains three-gluon vertices and ghost loops, and the dressed gluon now carries internal gluon segments bounded by the three-gluon vertices. The firing rule on these gluon segments is the pinch decomposition \eqref{eq:cornwall} together with paired ghost contributions, with dressed gluon propagators replacing the bare ones. The locality of gluon segments is preserved at all orders by the same recursive argument that preserves quark-segment locality, in that the pinch decomposition at order $L$ acts on objects already gauge-invariant at order $L-1$ and the BV bookkeeping ensures consistent ghost completion.

Third, the color content of $\hat\Sigma$ becomes richer at higher loops. The $C_F$-only simplification~\eqref{eq:CFinheritance} does not persist beyond one loop. At two loops the dressed internal gluon carries 3-gluon-vertex and ghost contributions through its self-energy, and these introduce $C_A$ and $T_F n_f$ Casimirs into $\hat\Sigma^{(2)}$ directly. The two-loop quark self-energy in QCD has color structures $C_F^2$, $C_F C_A$ and $C_F T_F n_f$, all of which contribute to the two-loop mass function. The Binosi-Papavassiliou two-loop result~\cite{Binosi:Papavassiliou:2002} establishes that $\hat\Sigma^{(2)}_{\rm QCD}$ retains all of these Casimir structures with the gauge-invariance property $\hat\Sigma^{(2)} = \Sigma_\xi^{(2)}\big|_{\xi=1}$.

A practical consequence is that the QED simplification that the entire one-loop mass function in QCD is $C_F$ times the abelian result is a one-loop feature, not an all-orders feature. At two loops and beyond, the mass function $m^{\text{ren},(L)}_{\rm QCD}(q)$ carries multi-Casimir structures that have no abelian analog, and the explicit non-abelian apparatus of Sections~\ref{sec:qcd}--\ref{sec:qcd_n2} becomes essential to the calculation.

\subsection{Summary}

The framework extends in two structurally independent directions, along the loop expansion via the exact WTI acting on dressed objects and across gauge groups via the pinch decomposition and the BV formalism. Both extensions preserve segment locality. The locality of the segment is a property of the algebraic identities (WTI on quark segments, pinch decomposition on gluon segments) and not of the loop count or of the gauge group. The complications that arise at higher loops or in non-abelian theories are richer within-segment structures and additional segment types, not failures of locality.

The one-loop QED calculation of Sections~\ref{sec:pinch}--\ref{sec:mass} is therefore the explicit demonstration of a mechanism that extends in both directions, to higher loop order by induction and to non-abelian gauge groups by the pinch decomposition and the BV formalism. The mass function $m(q)$ defined at one loop in QED is the first member of a family of gauge-invariant mass functions $m^{\text{ren},(L)}_G(q)$ indexed by loop order $L$ and gauge group $G$, each constructed by the same segment-local mechanism with appropriately dressed objects and appropriately generalized firing rules. The one-loop $L=1$, $G={\rm U}(1)$ case is the simplest member of this family and the one that admits the cleanest diagrammatic exposition; the higher members follow by inheritance.

\section{Discussion}
\label{sec:discussion}

The present approach differs from the standard PT in several respects. The standard PT works with on-shell external fermions and invokes the Dirac equation to identify the propagator-like pinch parts by their Dirac structure after an on-shell projection; the present approach identifies them intrinsically, through the segment-local WTI on the internal line, with no Dirac equation and no on-shell projection. The explicit Compton calculation keeps the external legs off-shell (Section~\ref{sec:explicit}) and cancels the boundary terms without the Dirac equation; for general multiplicity the two outermost legs may instead be taken on-shell without loss, since the self-energy depends only on the internal momentum. Process independence, which in the standard PT must be verified by explicit calculation, here follows automatically from the locality of the segment. The off-shell mass function $m(q)$, which in the standard PT framework can only be accessed indirectly through the BFM equivalence, emerges here as the primary output. In QED, the exact WTI suffices at all orders with no additional machinery (Section~\ref{sec:allorders}), whereas the standard PT in QCD requires the BV formalism. Both prescriptions agree on-shell, because the Dirac equation projects out exactly the boundary terms.

Placed among existing mass definitions, $m(q)$ fills a clear gap. The pole mass is gauge-invariant but defined only at $\slashed{q}=m$~\cite{Nielsen:1975,Breckenridge:1994gs,Gambino:1999ai,Kniehl:Sirlin:2008}, and the $\overline{\text{MS}}$ running mass $\bar{m}(\mu)$, though gauge-invariant, is a scheme-dependent parameter of the renormalization group rather than the mass in the dressed propagator~\cite{Tarrach:1981,Kronfeld:1998}. The momentum-dependent mass function has so far been available in fixed-gauge computations~\cite{Roberts:Schmidt:2020}; the present construction makes it gauge-invariant at every momentum and scheme-independent in on-shell parameters, reducing to the pole mass at $\slashed{q}=m$, and thereby extends the fermion mass from a single number to a fully determined function of the momentum. Its scalar companion $\hat M(q^2)$ of Section~\ref{sec:IR} is in addition infrared finite and renormalization-group invariant, so the comparison with Schwinger--Dyson and lattice mass functions is direct.

A notable consequence concerns the status of virtual particles. In the conventional formalism, the internal fermion line is not a well-defined object, since the propagator $\Sf(q) = i/(\slashed{q}-m_B-\Sigma_\xi(q))$ depends on the gauge parameter $\xi$, and any question about the virtual fermion (its mass, its spacetime behavior) is gauge-dependent and therefore not physically meaningful. The gauge-invariant propagator $\hat\Sf^{\text{ren}}(q) = i/(\slashed{q}-m(q))$ changes this, and the internal fermion line becomes a well-defined, gauge-invariant object. The mass function $m(q)$ is the most immediate output, but the gauge-invariant propagator itself is the more fundamental result. In this precise sense the virtual fermion is on the same footing as the on-shell one, its mass being gauge-invariant and, for the canonical reference $\xi_0=1$ and up to the infrared regulator of the residue subtraction (Section~\ref{sec:IR}), as well determined as the pole mass $m$, differing only in that it depends on the momentum flowing through the line. Gauge invariance is necessary but not sufficient for direct observability, as noted above, so the claim is not that the internal line is measured in isolation but that the distinction between real and virtual, as far as the mass is concerned, is kinematic rather than dynamical: it is the distinction between $m(m) = m$ and $m(q)$ at $\slashed{q}\neq m$, both values of the same gauge-invariant function.

This continuity has a deeper origin. In the LSZ formalism, the asymptotic particle is declared free, yet its mass $m$ is the pole of the \textit{fully dressed} propagator; it is entirely a product of interactions. The internal line in this framework carries the same dressed propagator $\hat\Sf^{\text{ren}}(q)$, with the same self-energy $\hat\Sigma^{\text{ren}}$, evaluated at a different momentum. The on-shell particle is not undressed and the off-shell particle is not differently dressed; they are the same dressed object, and $m(q)$ is the single function that describes the dressing at every momentum. There was never a dynamical boundary between real and virtual, only a kinematic one. At the pole the propagator diverges and the particle can propagate to asymptotic distances; away from the pole it cannot. But the mass, which encodes the dynamical content of the dressing, is continuous and gauge-invariant throughout.

\appendix
\section{The gauge-dependent part of the mass function}

The double application of the WTI in (\ref{eq:WTI_tree}) gives  \cite{Binosi:Papavassiliou:2002}
\begin{equation}
  \slashed\ell\Sf(q-\ell)\slashed\ell  = -\SfI(q)\Sf(q-\ell)\SfI(q)
    + 2\SfI(q) - \SfI(q-\ell).
  \label{eq:double}
\end{equation}
Upon integration, the first term becomes
\begin{equation}
- i (-ig)^2 C_2 \dL \frac{1}{\ell^4}\SfI(q)\Sf(q-\ell)\SfI(q) = g^2 C_2 (\slashed{q}-m) \dL \frac{1}{\ell^4(\slashed{q}-\slashed{\ell}-m)} (\slashed{q}-m).
\end{equation}
Reducing the enclosed Dirac structure
modulo $\slashed{q}^2=q^2$ yields a piece linear in $(\slashed{q}-m)$ and, for $m\neq0$, one proportional to $m\,(\slashed{q}-m)^2$.  The second term is
$\SfI(q)$ times a $q$-independent integral. The third
\begin{equation}
\begin{split}
- i (-ig)^2 C_2 \dL \frac{1}{\ell^4}\SfI(q-\ell) &=  g^2 C_2 \dL \frac{\slashed{q}-\slashed{\ell}-m}{\ell^4} \\
 &= g^2 C_2 (\slashed{q}-m) \dL \frac{1}{\ell^4}.
\end{split}
\end{equation}
separates into $(\slashed{q}-m)$ times a $q$-independent integral plus an odd integrand that vanishes by Lorentz symmetry. 

The result is the main-text form $\Sigma_P(q) = (1-\xi)\,(\slashed{q}-m)\,\Omega(q)$, with the Dirac-valued one-loop coefficient given explicitly by
\begin{equation}
  \Omega(q) = \frac{g^2 C_2}{16\pi^2}\,\bigl[\,B(q^2) + m\,(\slashed{q}-m)\,C(q^2)\,\bigr].
  \label{eq:Omega_explicit}
\end{equation}
Here $B(q^2)=B_0(q^2,0,m^2)+(q^2+m^2)\,C(q^2)$ collects the standard massless-boson, mass-$m$ fermion bubble $B_0$, which carries the ultraviolet divergence and the $q$-independent constants, and $C(q^2)$ is the finite coefficient of the quadratic piece. The latter is fixed by the rank-one doubled-propagator integral
\begin{equation}
  \dL \frac{\ell^\mu}{\ell^4\,[(q-\ell)^2-m^2]}
  = \frac{-i}{16\pi^2}\,q^\mu\,C(q^2),
  \label{eq:Cdef}
\end{equation}
which a single Feynman parameter and the standard momentum shift evaluate to
\begin{equation}
  C(q^2) = \int_0^1\! dx\,\frac{x}{m^2-x q^2}
  = -\frac{1}{q^2}-\frac{m^2}{q^4}\ln\!\Bigl(1-\frac{q^2}{m^2}\Bigr).
  \label{eq:Cexplicit}
\end{equation}
The factor $(\slashed{q}-m)$ vanishes on-shell $\slashed{q}=m$ but not in general, so the self-energy, and hence the propagator, is gauge-dependent off-shell. For massless fermions the quadratic coefficient $m\,C(q^2)$ vanishes through its explicit factor of $m$ (the function $C(q^2)\to-1/q^2$ itself stays finite), so $\Sigma_P\to(1-\xi)\frac{g^2 C_2}{16\pi^2}(\slashed q-m)B(q^2)$ is purely linear there. Only the $q^2$-dependent parts of $B$ and $C$ carry physical content; the $q^2$-independent constants drop out of every on-shell-subtracted difference.


\end{document}